\documentclass[12pt,a4paper]{amsart}
\usepackage[vmargin=72pt,hmargin=96pt]{geometry}
\usepackage[T1]{fontenc}
\usepackage[utf8]{inputenc}
\usepackage[spanish,english]{babel}
\usepackage{amsfonts,amsmath,amsthm}
\usepackage[all]{xy}
\usepackage{graphicx}
\usepackage{suffix}
\usepackage{color}
\newtheorem{theorem}{Theorem}[section]

\newtheorem{lemma}[theorem]{Lemma}
\newtheorem{proposition}[theorem]{Proposition}
\theoremstyle{definition}
\newtheorem{definition}[theorem]{Definition}

\theoremstyle{remark}
\newtheorem{example}[theorem]{Example}
\newtheorem{remark}[theorem]{Remark}

\numberwithin{equation}{section}

\newcommand{\rbra}{\left(}       \newcommand{\rket}{\right)}
       
\newcommand{\cbra}{\left\{}      \newcommand{\cket}{\right\}}
\newcommand{\abra}{\left\langle} \newcommand{\aket}{\right\rangle}
       
\newcommand{\HH}{\mathcal{H}}
\newcommand{\LL}{\mathcal{L}}
\newcommand{\MM}{\mathcal{M}}
\newcommand{\bbR}{\mathbb{R}}
\newcommand{\To}{\longrightarrow}
\newcommand{\Mapsto}{\longmapsto}
\newcommand{\tensor}{\otimes}
\newcommand{\Horizontal}{\mathrm{Hor}}
\newcommand{\horizontal}{\mathsf{h}}
\newcommand{\Vertical}{\mathrm{Vert}}

\newcommand{\Hor}{\Horizontal}
\newcommand{\hor}{\horizontal}
\renewcommand{\Vert}{\Vertical} 

\newcommand{\Smooth}{\mathcal{C}^\infty}
\newcommand{\Sections}{\mathrm{Sec}}
\newcommand{\Fields}{\mathfrak{X}}
\newcommand{\Forms}{\Omega}

\newcommand{\Affine}{\mathrm{Aff}}
\newcommand{\Aff}{\Affine}

\newcommand{\leg}{\mathop{\mathrm{leg}}\nolimits}
\newcommand{\Leg}{\mathop{\mathrm{Leg}}\nolimits}
\newcommand{\pairing}{\abra\,\cdot\,,\cdot\,\aket}
\newcommand{\diff}{d}
\newcommand{\dt}{\diff t}
\newcommand{\du}{\diff u}
\newcommand{\dx}{\diff x}

\newcommand{\dmx}{\diff^mx}
\newcommand{\dmpx}{\diff^{m+1}x}
\newcommand{\dmpxi}{\diff^m x_i}

\newcommand{\cota}{\mathbb{T}^*\tilde{E}}

\newcommand{\dmxi}{\diff^{m-1}x_i}

\newcommand{\dd}[2][{}]{\frac{\diff#1}{\diff#2}} 
\WithSuffix\newcommand\dd*[2][{}]{\diff#1/\diff#2} 

\newcommand{\vardd}[2][{}]{\frac{\delta#1}{\delta#2}} 
\WithSuffix\newcommand\vardd*[2][{}]{\delta#1/\delta#2} 
\newcommand{\pp}[2][{}]{\frac{\partial#1}{\partial#2}} 
\WithSuffix\newcommand\pp*[2][{}]{\partial#1/\partial#2} 
\newcommand{\ppt}{\pp[]t}
\WithSuffix\newcommand\ppt*{\pp*[]t} 
\newcommand{\FSection}[1]{\widetilde{#1}}

\newcommand{\ie}{\emph{i.e.}}
\newcommand{\refer}{\emph{ref.}}
\newcommand{\quand}{\quad\textrm{and}\quad}
\newcommand{\qquand}{\qquad\textrm{and}\qquad}

\newcommand{\st}{\textrm{ s.t. }}

\title[]{Hamilton-Jacobi theory in Cauchy data space}
\author[C. M. Campos]{Cédric M. Campos}
\address{C.M. Campos: Dept. Matem\'atica Aplicada and IMUVA, Fac. Ciencias, UVA,
				Paseo de Bel\'en 7, 
				47011 Valladolid, Spain}
\address{M. de Le\'on, D. mart\'in de Diego, M. Vaquero: Instituto de Ciencias Matemáticas (CSIC-UAM-UC3M-UCM),
         Calle Nicolás Cabrera 13--15,
         Campus Cantoblanco,
         28049 Madrid, Spain}	
\email{cedricmc@uva.es}
\author[M. de León]{Manuel de León}
\email{mdeleon@icmat.es}
\author[D. Martín de Diego]{David Martín de Diego}
\email{david.martin@icmat.es}
\author[M. Vaquero]{Miguel Vaquero}
\email{miguel.vaquero@icmat.es}
\thanks{\noindent This work has been partially supported by MINECO MTM 2013-42870-P, MICINN (Spain) MTM 2010-21186-C02-01, MTM 2009-08166-E, the European project IRSES-project ``Geomech-246981'' and the ICMAT Severo Ochoa project SEV-2011-0087. M. Vaquero wishes to thank MINECO for a FPI-PhD Position.}

\begin{document}

\begin{abstract}
Recently, M. de Le\'on {\it el al.} (\cite{LeMaMa09}) have developed a geometric Hamilton-Jacobi theory  for Classical Field Theories in the setting of multisymplectic geometry. Our purpose in the current paper is to establish the corresponding Hamilton-Jacobi theory in the Cauchy data space, and relate both approaches.
\end{abstract}
\maketitle
\tableofcontents

\section{Introduction}

Multisymplectic geometry is the natural arena to develop Classical Field Theories of first order. Indeed, a multisymplectic manifold is a natural extension of symplectic manifolds, and in addition, the canonical models for multisymplectic structures are just the bundles of forms on a manifold in the same vein that cotangent bundles provide the canonical models for symplectic manifolds. 

One can exploit this parallelism between Classical Mechanics and Classical Field Theories but one should go very carefully. Indeed, instead of a configuration manifold, we have now a configuration bundle $\pi : E \longrightarrow M$ such that its sections are the fields (the manifold $M$ represents the space-time manifold). The Lagrangian density depends on the space-time coordinates, the fields and its derivatives, so it is very natural to take the manifold of 1-jets of sections of $\pi$, $J^1\pi$, as the generalization of the tangent bundle in Classical Mechanics; then a Lagrangian density is a fibered mapping $\LL : J^1\pi \longrightarrow \Lambda^{m+1}M$ (we are assuming that $\dim M = m+1$). From the Lagrangian density one can construct the Poincar\'e-Cartan form which gives the evolution of the system.

On the other hand, the spaces of 1- and 2-horizontal $m+1$-forms on $E$ with respect to the projection $\pi$, denoted respectively by $\Lambda_1^{m+1} E$ and $\Lambda_2^{m+1} E$, are the arena where the Hamiltonian picture of the theory is developed. To be more precise, the phase space is just the quotient 
$$
\MM^o \pi = \Lambda^{m+1}_2 E/\Lambda^{m+1}_1 E
$$
and the Hamiltonian density is a section of $\Lambda^{m+1}_2E \longrightarrow \MM^o\pi$ (the Hamiltonian function appears when a volume form $\eta$ on $M$ is chosen, such that $\HH = H \, \eta$. The Hamiltonian section $\HH$ permits just to pull-back the canonical multisymplectic form of $\Lambda^{m+1}_2E$ to a multisymplectic form on $\MM^o\pi$.

Of course, both descriptions are related by the Legendre transform which send solutions of the Euler-Lagrange equations into solutions of the Hamilton equations. One important difference with the case of mechanics is that now we are dealing with partial differential equations and we lost in principle the integrability. In any case, the solutions in both sides are interpreted as integral sections of Ehresmann connections. For a detailed account of the multisymplectic formalism see the following references \cite{BiSniFi,CaIbLe99,LeMaSa04,LeMaMa96,LeMaMa09,LeMaSa04s,MuRo99,MuRo00a,GIMMsyI,GIMMsyII,Ro09,Sa04}.

The Hamilton-Jacobi problem for a Hamiltonian classical field theory given by a Hamiltonian $H$ consists in finding a family of functions $S^i = S^i(x^i, u^a)$ such that
\begin{equation}\label{hjf}
\frac{\partial S^i}{\partial x^i} + H(x^i, u^a, \frac{\partial S^i}{\partial u^a}) = f(x^i)
\end{equation}
for some function $f(x^i)$. Here  $(x^i, u^a)$ represent bundle coordinates in $E$.

In \cite{LeMaMa09} the authors have developed a geometric Hamilton-Jacobi theory in the context of multisymplectic manifolds. 
The procedure used there is an extension of that used in \cite{AbMa78}, but now considering Ehresmann connections instead of vector fields
as solutions of the Hamilton equations. Let us notice that an alternative approach to the Hamilton-Jacobi theory for Classical Field Theories has been developed in the context of $k$-symplectic and $k$-cosymplectic structures (see \cite{LeMaMaSaVi,LeSaVi,LeVi}).

One attempt to develop a combined formalism ($k$-symplectic and \linebreak $k$-cosymplectic) can be found in \cite{FoGo13,FoYe13}.

As it is well-know (see \cite{BiSniFi,GIMMsyII}) there is an alternative way to study Classical Field Theories, in an infinite dimensional setting. The idea is to split the space-time manifold $M$ in the space an time pieces. To do this, we need to take a Cauchy surface, that is, an $m$-dimensional submanifold $N$ of $M$ such that (at least locally) we have $M = \bbR \times N$. So, the space of embeddings from $N$ to $\MM^o \pi$ is known as the Cauchy space of data for a particular choice of a Cauchy surface. This allows us to integrate the multisymplectic form on $\MM^o \pi$ to the Cauchy data space and obtain a presymplectic infinite dimensional system, whose dynamics is related to the de Donder-Hamilton equations for $H$. 

The aim of the paper is to show how we can `ìntegrate" a solution of the Hamilton-Jacobi problem for $H$ in order to get a solution for the Hamilton-Jacobi problem for the infinite-dimensional presymplectic system.

\section{A multisymplectic point of view of Classical Field Theory}


We begin by briefly introducing the multisymplectic approach to Classical Field Theory: the Lagrangian setting and its Hamiltonian counterpart. The theory is set in a configuration fiber bundle, $E\rightarrow M$, whose sections represent the fields we are interested in. From a Lagrangian density defined on the first jet bundle of the fibration $\pi$, say $\LL:J^1\pi\rightarrow \Lambda^{m+1}M$ (we restrict ourselves to first order theories in this paper), we derive the Euler-Lagrange equations. On the Hamiltonian side, we start with a Hamiltonian density $\HH\colon J^1\pi^\dag\to\Lambda^{m+1}M$ to obtain Hamilton's equations. Here, $J^1\pi^\dag$ is the dual jet bundle, the field theoretic analogue of the cotangent bundle. The relation among these two settings is given, under proper regularity, by the Legendre transform. This classical introductory part can also be found in \cite{CaGuMa12,CaIbLe99,LeMaSa04,LeMaMa96}, besides, useful references are \cite{BiSniFi,Ca10,CaIbLe99,Sa04}.

From now on, $\pi\colon E\to M$ will always denote a fiber bundle of rank $n$ over an $(m+1)$--dimensional manifold, \ie\ $\dim M=m+1$ and $\dim E=m+1+n$. Fibered coordinates on $E$ will be denoted by $(x^i, u^\alpha)$, $0\leq i \leq m$, $1 \leq \alpha \leq n$; where $(x^i)$ are local coordinates on $M$. The shorthand $\dmpx = \dx^0\wedge\ldots\wedge\dx^m$ will represent the local volume form that $(x^i)$ defines and we will also use the notation $\dmpxi = i_{\pp*{x^i}}\dx^0\wedge\dx^1\wedge\ldots\wedge\dx^m$ for the contraction with the coordinate vector fields. Many bundles will be considered over $M$ and $E$, but all of them vectorial or affine. For these bundles, we will only consider natural coordinates. In general, indexes denoted with lower case Latin letters (resp. Greek letters) will range between 0 and $m$ (resp. 1 and $n$). The Einstein sum convention on repeated crossed indexes is always understood.

Furthermore, we assume $M$ to be orientable with fixed orientation, together with a determined volume form $\eta$. Its pullback to any bundle over $M$ will still be denoted $\eta$, as for instance $\pi^*\eta$. In addition, local coordinates on $M$ will be chosen compatible with $\eta$, which means such that $\eta=\dmpx$. To great extent, this form $\eta$ is not needed and our constructions can be generalized, although we are going to make use $\eta$ for the sake of simplicity.
\subsection{Multisymplectic structures} \label{sec:cft:multisymplectic:structures}
We begin reviewing the basic notions of multisymplectic geometry and present some examples.

\begin{definition} \label{def:multisymplectic.vector}
 Let $V$ denote a finite dimensional real vector space. A ($k+1$)--form $\Omega$ on $V$ is said to be \emph{multisymplectic} if it is non-degenerate, i.e., if the linear map
\begin{eqnarray*}
\flat_\Omega\colon
  V &\To    & \Lambda^kV^*\\
  v &\Mapsto& \flat_\Omega(v):=i_v\Omega
\end{eqnarray*}
is injective. In such a case, the pair $(V,\Omega)$ is said to be a \emph{multisymplectic} vector space of order $k+1$.
\end{definition}

\begin{definition}
A \emph{multisymplectic structure of order $k+1$} on a manifold $P$ is a closed ($k+1$)--form $\Omega$ on $P$ such that $(T_xP,\Omega(x))$ is multisymplectic for each $x\in P$. The pair $(P, \Omega)$ is called a \emph{multisymplectic manifold of order $k+1$}.
\end{definition}




The canonical example of a multisymplectic manifold is the bundle of forms over a manifold $N$, that is, the manifold $P=\Lambda^kN$.
\begin{example}[\textbf{The bundle of forms}] \label{ex:multisymplectic:canonical}
Let $N$ be a smooth manifold of dimension $n$, $\Lambda^kN$ be the bundle of $k$--forms on $N$ and $\nu\colon \Lambda^kN\to N$ be the canonical projection ($1\leq k\leq n$). The \emph{Liouville or tautological form of order $k$} is the $k$--form $\Theta$ over $\Lambda^kN$ given by
\[ \Theta(\omega)(X_1,\ldots,X_k) := \omega((T_\omega\nu)(X_1),\ldots, (T_\omega\nu)(X_k)), \]
for any $\omega\in\Lambda^kN$ and any $X_1,\ldots,X_k\in T_\omega(\Lambda^kN)$. Then, the \emph{canonical multisymplectic ($k+1$)--form} is
\[ \Omega := -\diff\Theta \,. \]

If $(x^i)$ are local coordinates on $N$ and $(x^i, p_{i_1\ldots i_k})$, with $1\leq i_1<\ldots<i_k\leq n$, are the corresponding induced coordinates on $\Lambda^kN$, then
\begin{gather}
\Theta = \sum_{i_1<\ldots<i_k}p_{i_1\ldots i_k}\dx^{i_1}\wedge\ldots\wedge\dx^{i_k} \,,
\intertext{and}
\Omega = \sum_{i_1<\ldots<i_k}-\diff p_{i_1\ldots i_k}\wedge\dx^{i_1}\wedge\ldots\wedge\dx^{i_k} \,.
\end{gather}
From this expression, it is immediate to check that $\Omega$ is indeed multisymplectic.
\end{example}

\begin{example}[\textbf{The bundle of horizontal forms}] \label{ex:multisymplectic:horizontal}
Let $\pi\colon E\to M$ be a fibration, that is, $\pi$ is a surjective submersion. Assume that $\dim M=m+1$ and $\dim E=m+1+n$. Given $1\leq r\leq n$, we consider the vector subbundle $\Lambda^k_rE$ of $\Lambda^kE$ whose fiber at a point $u \in E$ is the set of $k$--forms at $u$ that are $r$--horizontal with respect to $\pi$, that is, the set
\[ (\Lambda^k_rE)_u = \{ \omega\in\Lambda^k_uE \,:\, i_{v_r}\ldots i_{v_1}\omega=0 \quad \forall v_1,\ldots,v_r\in\Vert_u(\pi) \} \,, \]
where $\Vert_u(\pi)=\ker(T_u\pi)$ is the space of tangent vectors at $u\in E$ that are vertical with respect to $\pi$.

We denote by $\nu_r$, $\Theta_r$ and $\Omega_r$ the restriction to $\Lambda^k_rE$ of $\nu$, $\Theta$ and $\Omega$ respectively. It is easy to see that $(\Lambda^k_rE,\Omega_r)$ is a multisymplectic manifold. The case in which $r=1,2$ and $k=m+1$ are the interesting cases for multisymplectic field theory.

Let $(x^i,u^\alpha)$ denote adapted coordinates on $E$, , where $0\leq i \leq m$ and $1\leq \alpha \leq n$,  then they induce coordinates $(x^i, u^\alpha, p, p^i_\alpha)$ on $\Lambda^{m+1}_2E$ such that any element $\omega\in\Lambda^{m+1}_2E$ has the form $\omega=p\dmpx + p^i_\alpha\du^\alpha\wedge\dmpxi$, where $\dmpx = \dx^0\land ...\land \dx^m$ and $\dmpxi = i_{\pp{x^i}}\dmpx$. Therefore, we have that $\Theta_2$ and $\Omega_2$ are locally given by the expressions
\begin{gather}
\label{eq:form:tautological:coord}
\Theta_2 = p\dmpx + p^i_\alpha\du^\alpha\wedge\dmpxi \,,
\intertext{and}
\label{eq:form:mutlisymplectic:coord}
\Omega_2 = -\diff p\wedge\dmpx - \diff p^i_\alpha\wedge\du^\alpha\wedge\dmpxi \,.
\end{gather}

\end{example}

\subsection{Lagrangian formalism} \label{sec:cft:lagrangian:formalism}
As already mentioned, the Lagrangian formulation of Classical Field Theory is stated on the \emph{first jet manifold} $J^1\pi$ of the configuration bundle $\pi\colon E\to M$. This manifold is defined as the collection of tangent maps of local sections of $\pi$. More precisely,
\[ J^1\pi := \cbra T_x\phi \,:\, \phi\in\Sections_x(\pi),\ x\in M \cket \,. \]
The elements of $J^1\pi$ are denoted $j^1_x\phi$ and called the \emph{1st-jet of $\phi$ at $x$}.

 Adapted coordinates $(x^i, u^\alpha)$ on $E$ induce coordinates $(x^i, u^\alpha, u^\alpha_i)$ on $J^1\pi$ such that $u^\alpha_i(j^1_x\phi)=\pp*[\phi^\alpha]{x^i}|_x$. It is clear that $J^1\pi$ fibers over $E$ and $M$ through the canonical projections $\pi_{1,0}\colon J^1\pi \to E$ and $\pi_1\colon J^1\pi \to M$, respectively, and that $\pi_1=\pi\circ\pi_{1,0}$. In local coordinates, these projections are given by $\pi_{1,0}(x^i, u^\alpha, u^\alpha_i) = (x^i, u^\alpha)$ and $ \pi_1(x^i, u^\alpha, u^\alpha_i) = (x^i)$, notice that $\pi_1=\pi\circ\pi_{1,0}$.


Despite the conceptual similarities with the tangent bundle of a manifold, the first jet manifold is not a vector bundle but an affine one, which is a crucial difference. To be precise, the first jet manifold $J^1\pi$ is an affine bundle over $E$ modeled on the vector bundle $V(J^1\pi) = \pi^*(T^*M)\tensor_E\Vert(\pi)$.

\begin{remark}
For a review on affine bundles and its applications to Classical Field Theory see D.J. Saunders \cite{Sa}.
\end{remark}

The dynamics of a Lagrangian field system are governed by a \emph{Lagrangian density}, a fibered map $\LL\colon J^1\pi\to\Lambda^{m+1}M$ over $M$. The real valued function $L\colon J^1\pi \to \bbR$ that satisfies $\LL=L\eta$ is called the \emph{Lagrangian function}. Both Lagrangians permit to define the so-called \emph{Poincaré-Cartan forms}:
\begin{equation} \label{eq:pc-forms}
\Theta_\LL = L\eta + \abra S_\eta,\diff L\aket \in\Forms^{m+1}(J^1\pi) \quand \Omega_\LL= -\diff\Theta_\LL \in\Forms^{m+2}(J^1\pi) \,,
\end{equation}
where $S_\eta$ is a $(1,n)$ tensor field on $J^1\pi$ called \emph{vertical endomorphism} and whose local expression is
\begin{equation} \label{eq:vertical:endomorphism:eta}
S_\eta= (\du^\alpha-u^\alpha_j\dx^j)\wedge\dmxi \tensor \pp{u^\alpha_i} \,.
\end{equation}
In local coordinates, the Poincaré-Cartan forms read as follows
\begin{align}
\label{eq:pc-mform:coord}
\Theta_\LL =& \rbra L-u^\alpha_i\pp[L]{u^\alpha_i} \rket \dmpx + \pp[L]{u^\alpha_i}\du^\alpha\wedge \dmpxi \,, \\
\label{eq:pc-m1form:coord}
\Omega_\LL =& -(\du^\alpha-u^\alpha_j\dx^j) \wedge \rbra \pp[L]{u^\alpha}\dmpx - \diff\rbra\pp[L]{u^\alpha_i}\rket\wedge \dmpxi\rket \,.
\end{align}
Although we are using the volume form $\eta$ to define the Poincaré-Cartan forms, it can be easily seen that it only depends on the Lagrangian density. 

A \emph{critical point} of $\LL$ is a (local) section $\phi$ of $\pi$ such that
\[(j^1\phi)^*(i_X \Omega_\LL) = 0,\]
for any vector field $X$ on $J^1\pi$. A straightforward computation shows that this implies that
\begin{equation} \label{eq:euler-lagrange:coord}
(j^1\phi)^*\rbra \pp[L]{u^\alpha}-\dd{x^i}\pp[L]{u^\alpha_i}\rket = 0 \,,\ 1\leq \alpha\leq n \,.
\end{equation}
The above equations are called \emph{Euler-Lagrange equations}.

\begin{remark}
Indeed, it is enough to show that $(j^1\phi)^*(i_X \Omega_\LL) = 0$ holds for $\pi_1$-vertical vector fields, i.e. $X$ is a section of the natural projection $\Vert(\pi_1)\to J^1\pi$.
\end{remark}

\subsection{Hamiltonian formalism} \label{sec:cft:hamiltonian:formalism}
The dual formulation of the Lagrangian formalism is the Hamiltonian one, which is set in the affine dual bundles of $J^1\pi$. The \emph{(extended) affine dual bundle} $J^1\pi^\dag$ is the collection of real-valued affine maps defined on the fibers of $\pi_{1,0}:J^1\pi\to E$, namely
\[ (J^1\pi)^\dag := \Aff(J^1\pi,\bbR) = \cbra A\in\Aff(J^1_u\pi,\bbR) \,\colon\, u\in E \cket \,. \]
The \emph{(reduced) affine dual bundle} $J^1\pi^\circ$ is the quotient of $J^1\pi^\dag$ by constant affine maps, namely
\[ (J^1\pi)^\circ := \Aff(J^1\pi,\bbR)/\{f\colon E\to\bbR\} \,. \]
It is again clear that $J^1\pi^\dag$ and $J^1\pi^\circ$ are fiber bundles over $E$ but, in contrast to $J^1\pi$, they are vector bundles. Moreover, $J^1\pi^\dag$ is a principal $\bbR$--bundle over $J^1\pi^\circ$. The respective canonical projections are denoted $\pi_{1,0}^\dag\colon J^1\pi^\dag\to E$, $\pi_1^\dag=\pi\circ\pi_{1,0}^\dag$, $\pi_{1,0}^\circ\colon J^1\pi^\circ\to E$, $\pi_1^\circ=\pi\circ\pi_{1,0}^\circ$ and $\mu\colon J^1\pi^\dag\to J^1\pi^\circ$. The natural pairing between the elements of $J^1\pi^\dag$ and those of $J^1\pi$ will be denoted by the usual angular bracket,
\[ \pairing \colon J^1\pi^\dag\times_EJ^1\pi \To \bbR \,. \]
We note here that $J^1\pi^\circ$ is isomorphic to the dual bundle of $V(J^1\pi) = \pi^*(T^*M)\tensor_E\Vert(\pi)$.

Besides defining the affine duals of $J^1\pi$, we must also introduce the \emph{extended \emph{and} reduced multimomentum spaces}
\[ \MM\pi := \Lambda^{m+1}_2E \quand \MM^\circ\pi := \Lambda^{m+1}_2E/\Lambda^{m+1}_1E \,. \]
By definition, these spaces are vector bundles over $E$ and we denote their canonical projections $\nu\colon\MM\pi\to E$, $\nu^\circ\colon\MM^\circ\pi\to E$ and $\mu\colon\MM\pi\to\MM^\circ\pi$ (some abuse of notation here). Again, $\mu\colon\MM\pi\to\MM^\circ\pi$ is a principal $\bbR$--bundle. We recall from Example \ref{ex:multisymplectic:horizontal} that $\MM\pi$ has a canonical multisymplectic structure which we denote $\Omega$. On the contrary, $\MM^\circ\pi$ has not canonical multisymplectic structure, but $\Omega$ can still be pulled back by any section of $\mu\colon\MM\pi\to\MM^\circ\pi$ to give rise to a multisymplectic structure on $\MM^\circ\pi$.

An interesting and important fact is how the four bundles we have defined so far are related. We have that
\begin{equation} \label{eq:dual:identification}
J^1\pi^\dag \cong \MM\pi \quand J^1\pi^\circ \cong \MM^\circ\pi \,,
\end{equation}
although these isomorphisms depend on the base volume form $\eta$. In fact, the bundle isomorphism $\Psi:\MM\pi\to J^1\pi^\dag$ is characterized by the equation
\[ \abra\Psi(\omega),j^1_x\phi\aket\eta = \phi^*_x(\omega) \,,\quad \forall j^1_x\phi\in J^1_{\nu(\omega)}\pi \,,\quad \forall\omega\in \MM\pi \,. \]
We therefore identify $\MM\pi$ with $J^1\pi^\dag$ (and $\MM^\circ\pi$ with $J^1\pi^\circ$) and use this isomorphism to pullback the duality nature of $J^1\pi^\dag$ to $\MM\pi$.

\begin{remark}\label{rmk:identification:forms} We want to stress that
  the reader must keep in mind that $J^1\pi^\dag$ can be understood as
  a space of forms due to the previous identifications. This
  identification will be important in Theorem
  \ref{th:cft:hj:mutlisymplectic} and along the rest of this paper.
\end{remark}

Adapted coordinates in $\MM\pi$ (resp. $\MM^\circ\pi$) will be of the form $(x^i,u^\alpha,p,p^i_\alpha)$ (resp. $(x^i,u^\alpha,p^i_\alpha)$), such that
\[ p\dmpx+p^i_\alpha\du^\alpha\wedge\dmpxi\in\Lambda^{m+1}_2E \quad (pd^{m+1}x\in\Lambda^m_1E) \,. \]
Under these coordinates, the canonical projections have the expression
\[ \nu(x^i,u^\alpha,p,p^i_\alpha)=(x^i,u^\alpha) \,,\quad
   \nu^\circ(x^i,u^\alpha,p^i_\alpha)=(x^i,u^\alpha)\] 
	\begin{center} and \end{center}
\[
   \mu(x^i,u^\alpha,p,p^i_\alpha)=(x^i,u^\alpha,p^i_\alpha) \,; \]
and the above pairing takes the form
\[ \abra(x^i,u^\alpha,p,p^i_\alpha),(x^i,u^\alpha,u^\alpha_i)\aket = p + p^i_\alpha u^\alpha \,. \]
We also recall the local description of the canonical multisymplectic form $\Omega$ of $\MM\pi$,
\[ \Omega = -\diff p\wedge\dmpx - \diff p^i_\alpha\wedge\du^\alpha\wedge\dmpxi \,. \]

Now, we focus on the principal $\bbR$--bundle structure of $\mu\colon\MM\pi\to\MM^\circ\pi$. This structure arises from the $\bbR$--action
\begin{align*}
                \bbR\times\MM\pi &\,\To    \, \MM\pi\\
                     (t,\omega) &\,\Mapsto\, t\cdot\eta_{\nu(\omega)} + \omega \,.
\intertext{In coordinates,}
(t,(x^i,u^\alpha,p,p^i_\alpha)) &\,\Mapsto\, (x^i,u^\alpha,t+p,p^i_\alpha) \,.
\end{align*}
We will denote by $V_\mu\in\Fields(\MM\pi)$ the infinitesimal generator of the action of $\bbR$ on $\MM\pi$, which in coordinates is nothing else but $V_\mu=\pp{p}$. Since the orbits of this action are the fiber of $\mu$, then $V_\mu$ is also a generator of the vertical bundle $\Vert(\mu)$.

The dynamics of a Hamiltonian field system is governed by a \emph{Hamiltonian section}, say a section $h\colon\MM^\circ\pi\to\MM\pi$ of $\mu\colon\MM\pi\to\MM^\circ$. In presence of the base volume form $\eta$, the set of Hamiltonian sections $\Sections(\mu)$ is in one-to-one correspondence with the set of functions $\{\bar H\in\Smooth(\MM\pi) : V_\mu(\bar H)=1 \}$ and with the set of \emph{Hamiltonian densities}, that is, fibered maps $\HH\colon\MM\pi\to\Lambda^{m+1}M$ over $M$ such that $i_{V_\mu}\diff\HH=\eta$. Given a Hamiltonian section $h\colon\MM^\circ\pi\to\MM\pi$, the corresponding Hamiltonian density is
\[ \HH(\omega) = \omega - h(\mu(\omega)) \,,\ \forall\omega\in\MM\pi \,. \]
Conversely, given a Hamiltonian density $\HH\colon\MM\pi\to\Lambda^{m+1}M$, the corresponding Hamiltonian section is characterized by the condition
\[ \mathrm{im}h=\HH^{-1}(0) \,. \]
Obviously, $\HH=\bar H\eta$. In adapted coordinates,
\begin{align}
\label{eq:hamiltonian:section:coord}
h(x^i, u^\alpha, p^i_\alpha) &= (x^i, u^\alpha, p = -H(x^i, u^\alpha, p^i_\alpha), p^i_\alpha) \,,\\
\label{eq:hamiltonian:density:coord}
\HH(x^i, u^\alpha, p, p^i_\alpha) &= \rbra p + H(x^i, u^\alpha, p^i_\alpha)\rket\dmpx \,.
\end{align}
The locally defined function $H$ is called the \emph{Hamiltonian function} and it must not be confused with the globally defined mapping $\bar H$ such that $\HH=\bar H\eta$.

A \emph{critical point} of $\HH$ is a (local) section $\tau$ of $\pi\circ\nu\colon\MM\pi\to M$ that satisfies the \emph{(extended) Hamilton-De Donder-Weyl equation}
\begin{equation} \label{eq:hamilton-dedonder-weyl:ext}
\tau^*i_X(\Omega+\diff\HH) = 0 \,,
\end{equation}
for any vector field $X$ on $\MM\pi$.
A \emph{critical point} of $h$ is a (local) section $\tau$ of $\pi\circ\nu^\circ\colon\MM^\circ\pi\to M$ that satisfies the \emph{(reduced) Hamilton-De Donder-Weyl equation}
\begin{equation} \label{eq:hamilton-dedonder-weyl:red}
\tau^*(i_X\Omega_h) = 0 \,,
\end{equation}
for any vector field $X$ on $\MM^\circ\pi$ and where $\Omega_h=h^*(\Omega+\diff\HH)=h^*\Omega$.
\begin{remark}
Actually, it is enough to show that Equation \eqref{eq:hamilton-dedonder-weyl:ext} (resp. Equation \eqref{eq:hamilton-dedonder-weyl:red}) holds for vector fields that are vertical with respect to the bundle $\pi_1=\pi\circ\nu$ (resp. $\pi_1^\circ=\pi\circ\nu^\circ$).
\end{remark}
A straightforward but cumbersome computation shows that both, \eqref{eq:hamilton-dedonder-weyl:ext} and \eqref{eq:hamilton-dedonder-weyl:red}, are equivalent to the following set of local equations known as \emph{Hamilton's equations}:
\begin{equation} \label{eq:hamilton:coord:red}
\pp[\tau^\alpha]{x^i} = \pp[H]{p_\alpha^i} \circ \tau \,, \qquad
\pp[ \tau_\alpha^i]{x^i} = -\pp[H]{u^\alpha} \circ \tau \,,
\end{equation}
where $\tau^\alpha=u^\alpha\circ \tau$ and $\tau_\alpha^i=p_\alpha^i\circ\tau$.

We introduce now Ehresmann connections in order to write the
infinitesimal counterpart of the previous equations. Thus, an
Ehresmann connection on the bundle $\MM\pi^\circ\rightarrow M$ is
given by a distribution $\mathbf{H}$ in $T\MM\pi^\circ$ which is complementary
to the vertical one,
$\Vert(\pi\circ\nu^\circ)=\ker(T\pi\circ\nu^\circ)$. To see more
details about Eheresmann connections in this context see
\cite{LeMaSa04s,Sa04} and the Appendix A.
Let $\hor$ be the horizontal projector of an Ehresmann connection in the bundle $\pi_1^{\circ}$. We introduce now what should be understood as the infinitesimal description of the Hamilton's equations through the following proposition. 

\begin{proposition} \label{th:hamilton:connection}
If the horizontal projector $\hor$ of an Ehresmann connection satisfies
\begin{equation} \label{eq:hamilton:connection}
i_\hor\Omega_h=m\Omega_h \,.
\end{equation}
then any horizontal integral section $\sigma$ of the connection is a solution of Hamilton's equations.
\end{proposition}

An equivalent proposition could be set on the Lagrangian side using the form $\Omega_{\mathcal{L}}$.

\subsection{Equivalence between both formalisms} \label{sec:cft:equivalence}
In this section, we present the equivalence between the Lagrangian and Hamiltonian formalisms of classical field theories for the case when the Lagrangian function is (hyper)regular (see below).

Let $\LL$ be a Lagrangian density. The \emph{(extended) Legendre transform} is the bundle morphism $\Leg_\LL\colon J^1\pi\to\MM\pi$ over $E$ defined as follows:
\begin{equation} \label{eq:legendre.transform.ext}
\Leg_\LL(j^1_x\phi)(X_1,\ldots,X_m) := (\Theta_\LL)_{j^1_x\phi}(\widetilde{X}_1,\ldots,\widetilde{X}_m),
\end{equation}
for all $j^1_x\phi \in J^1\pi $ and $X_i \in T_{\phi(x)}E$, where $\widetilde{X}_i \in T_{j^1_x\phi}J^1\pi$ are such that $T\pi_{1,0}(\widetilde{X}_i)=X_i$. The \emph{(reduced) Legendre transform} is the composition of $\Leg_\LL$ with $\mu$, that is, the bundle morphism
\begin{equation} \label{eq:legendre.transform.red}
\leg_\LL := \mu \circ \Leg_\LL\colon J^1\pi \to \MM^\circ\pi \,.
\end{equation}
In local coordinates,
\begin{eqnarray}
\label{eq:legendre:transform:ext:coord}
\Leg_\LL(x^i, u^\alpha, u^\alpha_i) &=& \rbra x^i, u^\alpha, L-\pp[L]{u^\alpha_i}u^\alpha_i, \pp[L]{u^\alpha_i}\rket \,, \\
\label{eq:legendre:transform:red:coord}
\leg_\LL(x^i, u^\alpha, u^\alpha_i) &=& \rbra x^i, u^\alpha, \pp[L]{u^\alpha_i}\rket \,,
\end{eqnarray}
where $L$ is the Lagrangian function associated to $\LL$, \ie\ $\LL=L\eta$.

From the definitions, we deduce that $(\Leg_\LL)^*(\Theta)= \Theta_\LL, (\Leg_\LL)^*(\Omega) = \Omega_\LL$, where $\Theta$ is the Liouville $m$--form on $\MM\pi$ and $\Omega$ is the canonical multisymplectic ($m+1$)--form. In addition, we have that the Legendre transformation $\leg_\LL\colon J^1\pi\to \MM^\circ\pi$ is a local diffeomorphism, if and only if, the Lagrangian function $L$ is \emph{regular}, that is, the Hessian $\rbra\frac{\partial^2L}{\partial u^\alpha_i\partial u^\beta_j}\rket$ is a regular matrix. When $\leg_\LL\colon J^1\pi\to \MM^\circ\pi$ is a global diffeomorphism, we say that the Lagrangian $L$ is \emph{hyper-regular}. In this case, we may define the Hamiltonian section $h\colon \MM^\circ\pi\To \MM\pi$
\begin{equation} \label{eq:legendre:hamiltonian:section}
 h = \Leg_\LL\circ\leg_\LL^{-1} \,,
\end{equation}
whose associated Hamiltonian density is
\begin{equation} \label{eq:legendre:hamiltonian:density}
 \HH(\omega) = \abra\omega,\leg_\LL^{-1}(\mu(\omega))\aket\eta - (\LL\circ\leg_\LL^{-1})(\mu(\omega)) \,,\quad \forall\omega\in\MM\pi \,.
\end{equation}
In coordinates,
\begin{equation} \label{eq:legendre:hamiltonian:section:coord}
h(x^i, u^\alpha, p^i_\alpha) = (x^i, u^\alpha, L(x^i,u^\alpha,u^\alpha_i) - p^i_\alpha u^\alpha_i, p^i_\alpha) \,,
\end{equation}
where $u^\alpha_i=u^\alpha_i(\leg_\LL^{-1}(x^i, u^\alpha, p^i_\alpha))$. Accordingly,
\begin{equation} \label{eq:legendre:hamiltonian:density:coord}
\HH(x^i, u^\alpha, p, p^i_\alpha) = \rbra p + p^i_\alpha u^\alpha_i - L \rket\dmx
\end{equation}
and the Hamiltonian function is
\begin{equation} \label{eq:legendre:hamiltonian:function:coord}
H(x^i, u^\alpha, p^i_\alpha) = p^i_\alpha u^\alpha_i - L \,.
\end{equation}

\begin{theorem}[\refer\ \cite{LeMaMa96,MuRo00a,Ro09}]
Assume $\LL$ is a hyper-regular Lagrangian density. If $\phi$ is a solution of the Euler-Lagrange equations for $\LL$, then $\varphi=\leg_\LL\circ j^1\phi$ is a solution of the Hamilton's equations for $h$. Conversely, if $\varphi$ is a solution of the Hamilton's equations for $h$, then $\leg_\LL^{-1}\circ\,\varphi$ is of the form $j^1\phi$, where $\phi$ is a solution of the Euler-Lagrange equations for $\LL$.
\end{theorem}

From now on, we will assume every Lagrangian to be regular.

\section{Hamilton-Jacobi theory for multisymplectic systems} \label{sec:cft:hj:multisymplectic}
In this section we summarize the results given in \cite{LeMaMa09} about the Hamilton-Jacobi Theory in the multisymplectic setting in order to give the infinite dimensional counterpart in the next sections. The Hamiltonian formulation will be the only one used in this section. We start recalling the standard Hamilton-Jacobi theory from Classical Mechanics, useful references are \cite{AbMa78,Ar}.

Let $Q$ be the configuration manifold of a mechanical system and $T^*Q$ the corresponding phase space, which is equipped with the canonical symplectic form
\[ \omega_Q = \diff q^\alpha \wedge \diff p_\alpha \,, \]
where $(q^\alpha, p_\alpha)$ are natural coordinates in $T^*Q$. We denote $\pi_Q \colon T^*Q \to Q$ the canonical projection.

Let $H \colon T^*Q \To \bbR$ be a Hamiltonian function and $X_H$ the corresponding Hamiltonian vector field, that is, the one that satisfies
\[ i_{X_H}\omega_Q = \diff H \,. \]
The integral curves $(q^\alpha(t), p_\alpha(t))$ of $X_H$ satisfy the Hamilton's equations:
\[ \dd[q^\alpha]t = \pp[H]{p_\alpha} \qquand \dd[p_\alpha]t = -\pp[H]{q^\alpha} \,. \]

The following theorem gives the relation between the Hamilton-Jacobi equation and the solutions of Hamilton's equations.
\begin{theorem} \label{th:hj:coord}
Let $\lambda$ be a closed 1--form on $Q$. The following conditions are equivalent:
\begin{enumerate}
\item If $\sigma\colon I\to Q$ satisfies the equation
\[ \dd[q^\alpha]t = \pp[H]{p_\alpha}\circ\lambda \,, \]
then $\lambda\circ\sigma$ is a solution of the Hamilton's equations;
\item $\diff (H\circ \lambda)=0$.
\end{enumerate}
\end{theorem}

\begin{remark} Since $\lambda$ is closed, locally we have $\lambda=dS$
  for a function $S$ depending on the local coordinates
  $(q^\alpha)$. Then, the equation $\diff (H\circ \lambda)=0$ reads
  locally $d\Big(H(q^\alpha,\displaystyle\frac{\partial S}{\partial
    q^\alpha})\Big)=0$. Moreover, on each connected component, the
  previous equation becomes $H(q^\alpha,\displaystyle\frac{\partial
    S}{\partial q^\alpha})=E$, where $E$ is a real constant. The last
  formula is know as the Hamilton-Jacobi equation.
\end{remark}

We can give the following interpretation. Define on $Q$ the vector field
\[ X_H^\lambda = T\pi_Q\circ X_H\circ \lambda \]
whose construction is illustrated by the below diagram
\[ \xymatrix{
   T^*Q \ar[dd]^{\pi_Q} \ar[rrr]^{X_H} &&& T(T^*Q) \ar[dd]^{T\pi_Q} \\
   &&&\\
      Q \ar@/^1.4pc/[uu]^\lambda\ar[rrr]^{X_H^\lambda} &&& TQ
   } \]
We then have the intrinsic version of Theorem \ref{th:hj:coord}.
\begin{theorem}[ref. \cite{AbMa78}] \label{th:hj:intrinsic}
Let $\lambda$ be a closed $1$--form on $Q$. Then the conditions below are equivalent:
\begin{enumerate}
\item $X_H^\lambda$ and $X_H$ are $\lambda$-related;
\item $\diff (H\circ \lambda)=0$.
\end{enumerate}
\end{theorem}

In the Classical Field framework, the role of the Hamiltonian vector field $X_H$ is played by a solution $\hor$ of the field equation \eqref{eq:hamilton:connection}, while the role of the $1$--form $\lambda$ above is now played by $\gamma$, a $2$-semibasic $(m+1)$--form, otherwise a section of the bundle $\pi_{1,0}^\dag\colon J^1\pi^\dag \to E$. Following the previous idea, we project along $\gamma$ the Ehresmann connection on $J^1\pi_1^\circ\to M$ to an Ehresmann connection on $E\to M$ whose horizontal projector is
\begin{equation} \label{reduced:connection}
\begin{array}{rcl}
\hor^\gamma(e)\colon T_eE & \To & T_eE \\
X & \Mapsto & \hor^\gamma(e)(X)=T_f\pi_{1,0}^\circ(\hor(f)(Y)),
\end{array}
\end{equation}
where $f=(\mu\circ\gamma)(e)$ and $Y$ is any vector of $T_fJ^1\pi^\dag$ which projects onto $X$ by $T\pi_{1,0}^\circ$. Notice that the Ehresmann connection given by $\hor^\gamma$ plays the role of the  projected vector field $X_H^\lambda$ in mechanics. 
\[
\xymatrix{
J^1\pi^\circ \ar[dr]^{\pi^\circ_1}&& \textrm{Connection } h \textrm{ on } J^1\pi^\circ\rightarrow M\ar[dd]^{\textrm{using $\gamma$ induces}}\\
J^1\pi^\dagger\ar[u]^\mu \ar[d]& M \\
E\ar@/^/[u]^\gamma\ar[ur]^\pi &&\textrm{Connection } h^\gamma \textrm{ on } E\rightarrow M}
\]

\begin{remark} 
We remit the reader to Appendix A for a brief survey on Ehresmann connections.
\end{remark}

\begin{theorem}[ref. \cite{LeMaMa09}] \label{th:cft:hj:mutlisymplectic}
Assume that $\gamma$ is closed and that the induced connection on $E\to M$, $\hor^\gamma$, is flat. Then the following conditions are equivalent:
\begin{enumerate}
\item If $\sigma$ is an integral section of $\hor$ then $\mu\circ\gamma\circ\sigma$ is a solution of the Hamilton's equations.
\item The $(n+1)$--form $h\circ\mu\circ\gamma$ is closed.
\end{enumerate}
\end{theorem}

The condition $\diff (h\circ\mu\circ\gamma)=0$ which happens to be equivalent to $(\mu\circ\gamma)^*\Omega_h=0$, corresponds to the generalization to Classical Field Theory of the Hamilton-Jacobi equation. Therefore we will refer to a form $\gamma$ satisfying it as a \emph{solution of the Hamilton-Jacobi equation}.

\begin{remark}
It can be seen (\cite{LeMaMa09}) that if we assume that $\lambda=dS$,
where $S$ is a $1$-semibasic $m$-form, then in local coordinates the
equation $\diff (h\circ\mu\circ\gamma)=0$ is equivalent to
$\displaystyle\frac{\partial S^i}{\partial x^i}+H(x^i,u^\alpha,
\displaystyle\frac{\partial S^i}{\partial u^\alpha})=f(x^i)$, where
$f(x^i)$ is a function on $M$. This is the usual way to wite the
Hamilton-Jacobi equations for Classical Field Theory \cite{La70,Ru73}.
\end{remark} 

\section{The space of Cauchy data} \label{sec:cft:cauchy:data}
In this section we develop the infinite-dimensional formulation of Hamilton's equations in order to introduce the Hamilton-Jacobi theory in infinite dimensions in the next section. We start introducing some basic definitions, the main references for this section will be \cite{Sa04,Va07}.

\begin{definition}
We say that an $m$--dimensional, compact, oriented and embedded submanifold $\Sigma$ of the base manifold $M$ is a \emph{Cauchy surface}.
\end{definition}

We will assume that that $\Sigma$ is endowed with a volume form, $\eta_\Sigma$, such that
\[ \int_\Sigma \eta_\Sigma=1 \,. \]

\begin{definition} \label{def:slicing}
A \emph{slicing} of $M$ is a diffeomorphism between $M$ and $\bbR\times \Sigma$, say
\[ \chi_M\colon \bbR\times \Sigma\to M \,. \]
\end{definition}

Observe that for each fixed $t\in\bbR$, $\chi_M(t,\cdot)\colon \Sigma\to M$ defines an embedding
\[ \begin{array}{rccl}
(\chi_M)_t\colon & \Sigma & \To & M \\
& x &\Mapsto & (\chi_M)_t(x)=\chi_M(t,x) \,.
\end{array} \]
We denote by $\Sigma_t=\mathop{Im}((\chi_M)_t)$ the image of $\Sigma$ by $(\chi_M)_t$ and by $\FSection{M}$ the space of such embeddings
\[ \FSection{M}=\{(\chi_M)_t\textrm{ such that }t\in\bbR\} \,, \]
which happens to be equivalent to $\bbR$, $\FSection{M}\equiv \bbR$. Without loss of generality, we may assume that $\Sigma$ is given by one of these embeddings, \ie\ there exists $t_0$ such that $\Sigma=\Sigma_{t_0}$. We will also use $(\chi_M)_t$ to denote the restriction of this map to its image, which happens to be a diffeomorphism between $\Sigma$ and $\Sigma_t$.

The use of the slicing $\chi$ is, obviously, to split $M$ onto time plus space and, particularly, to outline a 1--dimensional direction, which may be recovered infinitesimally. Let $\ppt*$ denote the vector field on $\bbR\times\Sigma$ characterizing the time translations $(t,x)\mapsto(t+s,x)$.

\begin{definition} \label{def:generator}
The vector field $\xi_M=(\chi_M)_*(\ppt*)\in\Fields(M)$ is the \emph{(infinitesimal) generator} of $\chi_M$. Its dual counterpart $(\chi_M^{-1})^*(\dt)\in\Forms(M)$ will still be denoted $\dt$.
\end{definition}

Let $\pi\colon E\to M$ be any bundle, the set
\[ \FSection{E} = \{ \sigma\colon\Sigma\to E \textrm{ such that } \pi\circ\sigma=(\chi_M)_t \textrm{ for some } t\in\bbR \} \]
is called the \emph{space of $\chi$--sections} of $E$. Indeed, it is a line bundle
\[ \begin{array}{rccl}
     \tilde\pi\colon & \FSection{E} & \To     & \bbR \\
                     &       \sigma & \Mapsto & t \st \pi\circ\sigma=(\chi_M)_t \,.
\end{array} \]

Consequently, a section of $\pi:E\rightarrow M$ induces a section of
$\overline{\pi}:\tilde{E}\rightarrow\mathbb{R}$, and conversely. This
correspondence just relates  the finite (multisymplectic) picture and
the infinite (presymplectic) one for a Classical Field Theory.




\begin{remark}
In the same spirit of \cite{GIMMsyII}, we assume that these spaces of embeddings are topologized in a way that they become infinite-dimensional smooth manifolds (see \cite{KrMi91} as well). 
\end{remark}


\begin{remark}\label{rmk:composition}
Observe that we still have the previous bundle structures by composition, for instance, from the bundle $\pi_{1,0}^\circ\colon J^1\pi^\circ\to E$ we can construct the bundle
\[
\begin{array}{rccl}
\FSection{\pi_{1,0}^\circ}&\FSection{J^1\pi^\circ }&\To&\FSection{E}\\ \noalign{\medskip}
& \sigma_{J^1\pi^\circ }&\to& \FSection{\pi_{1,0}^\circ}(\sigma_{J^1\pi^\circ })=\pi_{1,0}^\circ\circ\sigma_{J^1\pi^\circ }.
\end{array}
\]
We will use this procedure and notation to construct new bundles in the infinite dimensional setting from the ones on the finite dimensional framework. For example, in the same fashion we have the bundles $\FSection{\pi}\colon \FSection{E}\to \bbR$ and $\FSection{\pi_1^\circ}:\FSection{J^1\pi^\circ }\to\bbR $ since $\FSection{M}\equiv \bbR$.
\end{remark}

Now we give a short description of {\it tangent vectors} and some forms on the manifolds of embeddings following
\cite{Sa04, Va07}. We give the description in the $\FSection{J^1\pi^\circ }$ case, which is going to play the main role in what follows, being the others analogous.

Consider a differentiable curve from an open real interval $c\colon( -\epsilon,\epsilon)\to \FSection{J^1\pi^\circ }$ where $\epsilon$ is a positive real number and such that $c(0)=\sigma_{J^1\pi^\circ }$. Computing $\displaystyle\frac{dc}{dt}(0)$ it is easy to see that a \emph{tangent vector}, $\FSection{X}$, at a point $\sigma_{J^1\pi^\circ }\in\FSection{J^1\pi^\circ} $ is given by a map $\FSection{X}\colon \Sigma\To TJ^1\pi^\circ$ such that the following diagram is commutative
\[
\xymatrix{
\Sigma \ar[rr]^{\FSection{X}} \ar[rrd]_{\sigma_{J^1\pi^\circ }}&&  TJ^1\pi^\circ \ar[d]^{\tau_{J^1\pi^\circ}}\\
&&  J^1\pi^\circ
}
\]
and such that there exists a constant $k\in\bbR$ in a way that
\[T\pi_1^\circ(\FSection{X}(p))=k\xi_M(\pi_1^\circ(\sigma_{J^1\pi^\circ }(p))), \textrm{ for all } p\in\Sigma,\]
where we recall that $\xi_M$ denotes the generator introduced in \ref{def:generator} and $\tau_{J^1\pi^\circ}$ is the natural projection from the tangent bundle.

We show now how to construct {\it forms on the infinite dimensional
setting from forms on the finite dimensional side}. Once more, we give the description in the $\FSection{J^1\pi^\circ }$ case being the others analogous.

Let $\alpha$ be a $(k+m)$--form on $J^1\pi^\circ$, we define
the $k$--form $\mathbf{\FSection{\alpha}}$ on $\FSection{J^1\pi^\circ }$, such that for a point $\sigma_{J^1\pi^\circ}\in \FSection{J^1\pi^\circ }$ and $k$ tangent vectors $\FSection{X}_i\in T_{\sigma_{J^1\pi^\circ}}\FSection{J^1\pi^\circ }$ the pairing is given by
\begin{equation}\label{def:form}
\FSection{\alpha}(\sigma_{J^1\pi^\circ})(\FSection{X}_1,\ldots,\FSection{X}_k)=\int_\Sigma\sigma_{J^1\pi^\circ}^*(i_{\FSection{X}_1,\ldots,\FSection{X}_k}\alpha)
\end{equation}

The next lemma will be useful in this paper.
\begin{lemma}(ref. \cite{LeMaSa04s,Sa04})\label{lemma:differential}
Let $\alpha$ be a $k+m$--form, then $d(\FSection{\alpha})=\FSection{d\alpha}$.
\end{lemma}

\begin{remark}
The $2$--form $\FSection{\Omega}_h$, made out of $\Omega_h$ by this procedure, will play an important role describing the solutions of Hamilton's equations as infinite dimensional dynamical system.
\end{remark}

\begin{remark}\label{rmk:dt} The form $\FSection{dt\wedge \eta_{\Sigma}}$ is equal to $(\FSection{\pi_1^\circ})^*dt$.
\end{remark}

Now, we introduce local coordinates on the manifolds of embeddings using coordinates adapted to the slicing on $M$ as announced in the end of Section \ref{sec:cft:hamiltonian:formalism}. Let us work in coordinates adapted to the slicing on $M$, i.e., $(x^0,x^1,\ldots,x^n)$ are such that locally the $\Sigma_t$ are given by the level sets of the function $x^0$, moreover, we assume that in this coordinates the generator vector field $\xi_M$ is given by $\pp[]{x^0}$, which can be always achieved re-scaling the variable $x^0$. Actually we can assume that the coordinate $x^0$ is given by the function $t$ under the identification $\chi_M\colon\bbR\times \Sigma\to M $, where
\[
\begin{array}{rccl}
t:&\bbR\times \Sigma&\To&\bbR\\ \noalign{\medskip}
&(t_0,p)&\to& t(t_0,p)=t_0.
\end{array}
\]
Thus, from now on, making some abuse of notation (we are using $t$ to denote $t\circ \chi_M^{-1}$) we will take for granted that we are working with coordinates $(t,x^1,\ldots,x^n)$ as described before from now on. We want to explain that the choice of this coordinate $t$ is by no means arbitrary, it suggest the existence of a time parameter and so the generator vector field $\xi$ a time evolution direction. This is motivated by what happens for instance in Relativity (see \cite{GIMMsyI,GIMMsyII}).

Now, choosing coordinates adapted to the
fibration and to the base coordinates $(t,x^i)$, $1\leq i\leq m$ (adapted to the slicing), say $(t,x^i,u^\alpha)$,
 a point on $\FSection{E}$ is given by specifying functions
$u^\alpha(\cdot{})$ that depend on the coordinates on $\Sigma_t$,
i.e. $(x^i)$, $i=1,\ldots,n$. So ``coordinates'' on $\FSection{E}$ are
given by
\begin{equation}\label{coordinates:base}
(t,u^\alpha(\cdot{})), \quad t\in\bbR, \ u^\alpha=u^\alpha(x^1,\ldots,x^n)
\end{equation}
where the functions $u^\alpha$ belong to the chosen functional space.

\begin{remark}
Let us notice that this construction does not provide true local coordinates, but it is a nice way to determine elements of these different spaces of mappings.
\end{remark}

In the same way, choosing coordinates adapted to the slicing on $M$ as above and to the bundles
$J^1\pi^\circ\to M$ and $J^1\pi^\dag\to M$, say $(t,x^i,p^t_\alpha,p^i_\alpha)$ and  $(t,x^i,p,p^t_\alpha,\newline p^i_\alpha)$, defined by
\[
\begin{array}{rccl}
(x^i,u^\alpha,p^t_\alpha,p^i_\alpha)&\to
&[p^t_\alpha du^\alpha \wedge \dmx-p^i_\alpha du^\alpha\wedge dt\wedge \dmxi]\in \MM\pi^\circ_{| (x^i,u^\alpha)}.
\end{array}
\]
and
\[
\begin{array}{ll}
(x^i,u^\alpha,p,p^t_\alpha,p_\alpha^i)\to & pdt\wedge \dmx+p^t_\alpha du^\alpha \wedge \dmx \\\noalign{\medskip} &-p^i_\alpha du^\alpha\wedge dt\wedge \dmxi\in \MM\pi_{|(x^i,u^\alpha)}
\end{array}
\]
respectively, where $[\cdot{ }]$ denotes the equivalence class in the quotient and are using that $\dmx=dx^1\wedge dx^2\ldots\wedge dx^m$ and $\dmxi=i_{\pp[]{x^i}}\dmx$. There is certain abuse of notation here, but there is no room for confusion. Notice that we are using the identifications $J^1\pi^\dag\equiv \MM\pi$ and $J^1\pi^\circ\equiv \MM\pi^\circ$ again.

From the above discussion we deduce that, the points of $\FSection{J^1\pi^\circ}$ and $\FSection{J^1\pi^\dag }$ are
given by specifying respectively functions $p_\alpha^t(\cdot{}), \ p_
{\alpha}^i(\cdot{})$, and $p(\cdot{}), \ p_\alpha^t(\cdot{}), \ p_
{\alpha}^i(\cdot{})$ that depend on $(x^i)$, following the same construction that we introduced in the $\FSection{E}$ case. Thus, local coordinates on $\FSection{J^1\pi^\circ }$ are given by
\begin{equation}\label{coordinates:reduced}
(t,u^\alpha(\cdot{}),p_\alpha^t(\cdot{}),  p_
{\alpha}^i(\cdot{}))
\end{equation}
where $u^\alpha=u^\alpha(x^1,\ldots,x^n)$, $p_\alpha^t= p_\alpha^t(x^1,\ldots,x^n)$ and $p_
{\alpha}^i=p_
{\alpha}^i(x^1,\ldots,x^n)$.
Analogously, coordinates on $\FSection{J^1\pi^\dag }$ can be given by
\begin{equation}\label{coordinates:extended}
(t,u^\alpha(\cdot{}),p(\cdot{}),p_\alpha^t(\cdot{}), p_
{\alpha}^i(\cdot{})).
\end{equation}

By the previous constructions we can consider the manifold
$\FSection{J^1\pi^\circ }$ endowed with the form
$\FSection{\Omega}_h$ obtained by the construction outlined in \ref{def:form}. Using Lemma \ref{lemma:differential} we obtain that $(\FSection{J^1\pi^\circ},\FSection{\Omega_h})$ is a presymplectic manifold. It is worth to notice that there is an obvious bijective correspondence between sections of the bundle $\FSection{\pi_1^\circ}\colon \FSection{J^1\pi^\circ }\to \bbR$ and sections of the bundle $\pi_1^{\circ}\colon J^1\pi^\circ\to M$. Given $\sigma$ a section of the bundle $\pi_1^\circ$ consider the section of $\FSection{\pi_1^\circ}$ given by $c(t)=\sigma_{|\Sigma_t}\circ (\chi_M)_t\in \FSection{J^1\pi^\circ}$. Conversely given $c$ a section of $\FSection{\pi_1^\circ}$ using the slicing $(\chi_M)_t$ in the obvious way we can construct a section of $\pi_1^\circ$.

The following theorem
allows us to interpret Hamilton's equations as an infinite dimensional
dynamical system

\begin{proposition}[ref. \cite{LeMaSa04s,Sa04,Va07}]\label{th:correspondence} A section $\sigma$ of $\pi_1^\circ$
 satisfies Hamilton's equations if and only if the corresponding curve
 $c(t)$ verifies
\[
i_{\dot{c}(t)}\FSection{\Omega}_h=0
\]
where $\dot{c}(t)$ denotes the time derivative of the curve.
\end{proposition}

\begin{remark}
One could easily check that since $c$ is a section of $\FSection{\pi_1^\circ}$ and due to Remark \ref{rmk:dt}, then $\FSection{dt\wedge \eta_{\Sigma}}(\dot{c}(t))=1$.
\end{remark}

\section{Hamilton-Jacobi theory on the space of Cauchy data}\label{sec:cft:hamilton-jacobi:cauchy}

Assume now that we have a solution $\gamma$ of the
Hamilton-Jacobi equation as stated in Section \ref{sec:cft:hj:multisymplectic} and a
connection $\hor$ on the bundle $\pi_1^\circ$ satisfying the field
equations \eqref{eq:hamilton:connection} and consider the reduced
connection $\hor^\gamma$ on the bundle $\pi$ constructed in
\ref{reduced:connection}. We show in this section how to induce a
solution of the Hamilton-Jacobi equation in the infinite dimensional
setting as well as the meaning of the Hamilton-Jacobi problem in this setting. 

Following the previous section we can induce a section of the bundle $\FSection{\pi_{1,0}^\circ}\colon \FSection{J^1\pi^\circ }\to \FSection{E}$ by
\[
\begin{array}{rccl}
\FSection{\gamma}\colon &\FSection{E}& \To & \FSection{J^1\pi^\circ } \\ \noalign{\medskip}
&\sigma_E& \to &\FSection{\gamma}(\sigma_E)=\mu\circ\gamma\circ\sigma_E.
\end{array}
\]

On the other hand we can induce vector fields $\FSection{X}^\hor$ and $\FSection{X}^{\hor^\gamma}$ from the connections $\hor$ and $\hor^\gamma$ by
\[
\begin{array}{rcccccl}
\FSection{X}^\hor\colon &\FSection{J^1\pi^\circ }&\To &T\FSection{J^1\pi^\circ }\\ \noalign{\medskip}
&\sigma_{J^1\pi^\circ} &\to &           \FSection{X}^\hor(\sigma)&\colon \Sigma &\to &TJ^1\pi^\circ\\ \noalign{\medskip}
       &&&                                             &p&\to &\FSection{X}^\hor(\sigma)(p)=\textrm{Hor}(\xi((\chi_M)_t(p)))(\sigma_{J^1\pi^\circ}(p)),
\end{array}
\]
where $\Hor(X)(y)$ represents the horizontal lift of the tangent vector $X$ to the point $y$.

In the same way we can construct the vector field $\FSection{X}^{\hor^\gamma}$ on $\FSection{E}$ using the horizontal lift of the connection $\hor^\gamma$.

\begin{remark} Notice that the vector field $\FSection{X}^{\hor^\gamma}$ just described can also be defined as the $\FSection{\gamma}$-projection of the vector field $\FSection{X}^\hor$, i.e. we have
\[
\FSection{X}^{\hor^\gamma}(\sigma_E)=T\FSection{\pi_{1,0}^\circ}(\FSection{X}^\hor(\FSection{\gamma}(\sigma_E
))) \textrm{, where } \sigma_E\in \FSection{E}.
\]
\end{remark}

In local coordinates, assuming that
\begin{equation}\label{gamma}
\gamma(t,x^i,u^\alpha)=(t,x^i,u^\alpha,\gamma_p(t,x^i,u^\alpha),\gamma_{p_\alpha^t}(t,x^i,
u^\alpha),\gamma_{p_\alpha^i}(t,x^i, u^\alpha))
\end{equation}
and using the following notation in local coordinates \[\FSection{\gamma}(t,\sigma_E^\alpha(\cdot{}))=(t,\cdot{},\sigma_E^\alpha(\cdot{}),\gamma_{p_\alpha^t}(t,x^i,
\sigma_E^\alpha(\cdot{})),\gamma_{p_\alpha^i}(t,x^i, \sigma_E^\alpha(\cdot{})))\] where $\sigma_E^\alpha=u^\alpha\circ\sigma_E$ for $\sigma_E\in\FSection{E}$, the expressions $\FSection{X}^\hor(\FSection{\gamma}(\sigma_E))$ and $\FSection{X}^{\hor^\gamma}(\sigma_E)$ become
\begin{equation}
\begin{array}{ll}
\FSection{X}^\hor(\FSection{\gamma}(t,\sigma_E^\alpha(\cdot{})))=&\displaystyle\ppt
+\Gamma_\alpha^0(\FSection{\gamma}(t,\sigma_E^\alpha(\cdot{})))\displaystyle\pp[]{u^\alpha}
+(\Gamma_0)_\alpha^0(\FSection{\gamma}(t,\sigma_E^\alpha(\cdot{})))\displaystyle\pp[]{p_\alpha^t}
\\ \noalign{\medskip}

&+(\Gamma_0)_\alpha^i(\FSection{\gamma}(t,\sigma_E^\alpha(\cdot{})))\displaystyle\pp[]{p_\alpha^i}
\end{array}
\end{equation}
and
\begin{equation}
\begin{array}{l}
\FSection{X}^{\hor^\gamma}(t,\sigma_E^\alpha(\cdot{}))=\displaystyle\ppt
+\Gamma_\alpha^0(\FSection{\gamma}(t,\sigma_E^\alpha(\cdot{})))\displaystyle\pp[]{u^\alpha}.
\end{array}
\end{equation}

Notice that for each $t_0\in\bbR$ we have the bundle given by restriction
\begin{equation}\label{definition:restriction}\pi_{t}\colon E_t\to \Sigma_{t}\end{equation}
where $\pi_{t}=\pi_{|\Sigma_{t}}$ and $E_t=E_{|\Sigma_{t}}$. This bundle will play an important role in the next section. Observe that the space of sections $\Gamma(\pi_t)$ is just the fiber $\FSection{\pi}^{-1}(t)$.

\begin{definition}
 For each of these bundles we can induce the \emph{restricted connection}, $\hor^\gamma_t$ in the obvious way, i.e., the horizontal projector of the restricted connection is given by the restriction of the horizontal projector of the connection $\hor^{\gamma}$.
\end{definition}

Now we introduce one of the main results of the paper, that is,
$\FSection{\gamma}$ is a solution of the Hamilton-Jacobi
equation. This means that $\FSection{\gamma}^*\FSection{\Omega}_h=0$
and in addition for any point $\sigma_E\in \FSection{E}$ which is an
integral manifold of the corresponding restricted connection we have
that $T\FSection{\gamma}(\sigma_E)(\FSection{X}^{\hor^\gamma})$
satisfies
$i_{T\FSection{\gamma}(\sigma_E)(\FSection{X}^{\hor^\gamma})}\FSection{\Omega}_h=0$.

\begin{remark}
In the Hamilton--Jacobi theory on classical Hamiltonian systems $(T^*Q, \omega_Q, H)$, a solution of the Hamilton-Jacobi problem is a (closed) section 
$\gamma : Q \longrightarrow T^*Q$ of $\pi_Q : T^*Q \longrightarrow Q$ (i.e. a closed 1-form on $Q$)
such that $H \circ \gamma = const$. But $d\gamma = 0$ iff $\gamma^* \omega_Q = 0$, because the last equation just means that $\gamma(Q)$ is a lagrangian submanifold of $(T^*Q, \omega_Q)$.
This fact justifies the chosen notion of solution for the Hamilton-Jacobi problem
in the current context. In the next section this notion
will be more clear.
\end{remark}

\begin{theorem}\label{th:hamilton:jacobi:infinity}
The section $\FSection{\gamma}$ satisfies:
\begin{enumerate}
\item $\FSection{\gamma}^*\FSection{\Omega}_h=0$.
\item $i_{T\FSection{\gamma}(\sigma_E)(\FSection{X}^{\hor^\gamma})}\FSection{\Omega}_h=0$ for all $\sigma_E\in \FSection{E}$ which is an integral submanifold of the connection $\hor^\gamma_{\FSection{\pi}(\sigma_E)}$.
\end{enumerate}
\end{theorem}

\proof
The proof $i)$ is a direct consequence of the results in \cite{LeMaSa04s,Sa04}. In particular it is shown there that $\FSection{\gamma}^*\FSection{\Omega}_h=\widetilde{(\gamma^*\Omega_h)}$ and so the result holds recalling that $\gamma^*\Omega_h=d(h\circ\mu\circ\gamma)=0$ and $\gamma$ is a solution of the Hamilton-Jacobi equation.

The proof of $ii)$ will be done in local coordinates. Assume that
$\gamma$ is given by expression \ref{gamma}, i.e.
\[
\gamma(t,x^i,u^\alpha)=(t,x^i,u^\alpha,\gamma_p(t,x^i,u^\alpha),\gamma_{p_\alpha^t}(t,x^i,
u^\alpha),\gamma_{p_\alpha^i}(t,x^i, u^\alpha))
\]
and $u^\alpha\circ\sigma_E=\sigma_E^\alpha$ where $\sigma_E^\alpha$. So
\[
\begin{array}{ll}
T\FSection{\gamma}(\FSection{X}(\sigma_E))&=\displaystyle\ppt+\Gamma_0^\alpha(\FSection{\gamma}(t,\sigma_E^\alpha(\cdot{})))\displaystyle\pp[]{u^\alpha}
+\Big(\displaystyle\pp[\gamma_\alpha^t]{t}+\displaystyle\pp[\gamma_\alpha^t]{u^\beta}\Gamma_0^\beta(\FSection{\gamma}(t,\sigma_E^\alpha(\cdot{})))\Big)\displaystyle \pp[]{p_\alpha^t}
\\ \noalign{\medskip}
&+\Big(\displaystyle\pp[\gamma_\alpha^i]{t}+\displaystyle\pp[\gamma_ {\alpha}^i]{u^\beta}\Gamma_0^\beta(\FSection{\gamma}(t,\sigma_E^\alpha(\cdot{})))\Big)\displaystyle \pp[]{p_\alpha^i}.
\end{array}
\]

An easy computation shows that for any $\FSection{\pi_{1}^\circ}$-vertical tangent vector $\FSection{\xi}\in T_{\FSection{\gamma}(\sigma_E)}\FSection{J^1\pi^\circ}$ \[\FSection{\xi}=\xi_{u^\alpha}(\cdot{})\pp[]{ u^\alpha}+\xi_{p_\alpha^t}(\cdot{})\pp[]{p_\alpha^t}+\xi_{p_\alpha^i}(\cdot{})\pp[]{p_\alpha^i},\]
by the definition of $\FSection{\Omega}_h$, the expression $\FSection{\Omega}_h(T\FSection{\gamma}(\sigma_E)(\FSection{X}^{\hor^\gamma}),\FSection{\xi})$ reads
\[
\begin{array}{ll}
\FSection{\Omega}_h(T\FSection{\gamma}(\sigma_E)(\FSection{X}^{\hor^\gamma}),\FSection{\xi})&=
\displaystyle\int_\Sigma \Big(\xi_{u^\alpha} (-\displaystyle\pp[\gamma_\alpha^i]{x^i}-\pp[\gamma_
{\alpha}^i]{u^\beta}\pp[\sigma_E^\beta]{x^i} -\pp[\gamma_
{\alpha}^0]{t}-\pp[\gamma_
{\alpha}^0]{u^\beta}\Gamma_0^\beta -\pp[H]{u^\alpha} )
\\ \noalign{\medskip}

&\displaystyle+\xi_{p_\alpha^i} ( \pp[\sigma_E^\alpha]{x^i}
-\pp[H]{p_\alpha^i} )


+\xi_{p_\alpha^t} (-\displaystyle\pp[H]{p_\alpha^t}+\Gamma_0^\alpha )\Big)\dmx.
\end{array}
\]

The second and the third terms vanish because $\sigma_E$ is an integral submanifold of the restricted connection and because $\displaystyle\Gamma_0^\alpha =\pp[H]{p^t_\alpha}$. In order to show that the second term vanishes we need to use the following lemma from \cite{LeMaMa09}.

\begin{lemma}[ref. \cite{LeMaMa09}] If $d\gamma=0$, then the following assertions are
 equivalent
\begin{enumerate}
\item $d(h\circ\mu\circ\gamma)=0$
\item $\displaystyle\pp[H]{u^\alpha}
+\displaystyle\pp[H]{p^i_\beta}\displaystyle\pp[\gamma_\beta^i]{u^\alpha}
+\displaystyle\pp[H]{p_\beta^0}\pp[\gamma_\beta^0]{u^\alpha}
+\displaystyle\pp[\gamma_\alpha^i]{x^i}
+\pp[\gamma_\alpha^0]{t}=0$
\end{enumerate}
\end{lemma}

Going back to the proof of Theorem \ref{th:hamilton:jacobi:infinity} recalling that
$\displaystyle\pp[\sigma_E^\alpha]{x^i}=\displaystyle\pp[H]{p^i_ {\alpha}}$ and $\Gamma_0^\alpha=\displaystyle\pp[H]{p^0_\alpha}$
and that since $d\gamma=0$ then the following equalities hold
\[
\begin{array}{lr}
\displaystyle\pp[\gamma_\alpha^0]{u^\beta}=\displaystyle\pp[\gamma_\beta^0]{y^\alpha}; &\displaystyle\pp[\gamma_\alpha^i]{u^\beta}=\displaystyle\pp[\gamma^i_\beta]{u^\alpha}
\end{array}
\]
and thus we have
\[
\begin{array}{l}
-\displaystyle\pp[\gamma_\alpha^i]{x^i}-\displaystyle\pp[\gamma_
{\alpha}^i]{u^\beta}\displaystyle\pp[\sigma_E^\beta]{x^i} -\displaystyle\pp[\gamma_
{\alpha}^0]{t}-\displaystyle\pp[\gamma_
{\alpha}^0]{u^\beta}\Gamma_0^\beta -\displaystyle\pp[H]{u^\alpha}
\\ \noalign{\medskip}
=-\displaystyle\pp[\gamma_\alpha^i]{x^i}-\displaystyle\pp[\gamma_
{\beta}^i]{u^\alpha}\displaystyle\pp[H]{p_\beta^i} -\displaystyle\pp[\gamma_
{\alpha}^0]{t}-\displaystyle\pp[\gamma_
{\beta}^0]{u^\alpha}\displaystyle\pp[H]{p^i_ {\alpha}} -\displaystyle\pp[H]{u^\alpha}
\\ \noalign{\medskip}
=\displaystyle\pp[H]{u^\alpha}-\displaystyle\pp[H]{u^\alpha}=0
\end{array}
\]

Using the fact that in order to see that $i_{T\FSection{\gamma}(\sigma_E)(\FSection{X}^{\hor^\gamma})}\FSection{\Omega}_h=0$ is enough to prove it for $\FSection{\pi_{1}^\circ}$-vertical tangent vector the result follows.
\qed

\section{The space $\mathbb{T}^*\tilde{E}$ ($T^*\mathcal{E}^\tau$)}
In this section we are going to introduce the phase space $\cota$, which in the terminology of \cite{GIMMsyI} is the space denoted by ``$T^*\mathcal{E}^t$''. In order to do that, we have to start with a Lagrangian density.

Recall that $\pi\colon E\to M$ denotes a fiber bundle of rank $n$ over an $(m+1)$--dimensional manifold and $J^1\pi$ its first jet bundle, where we assume a Lagrangian density is given $\mathcal{L}:J^1E\to \Lambda^{m+1}M$. The submanifold $\Sigma$ is endowed with a volume form $\eta_\Sigma$.

From now on, we assume that we have a slicing $\chi_M$ on the manifold $M$. We will also assume that we have a compatible slicing.
\begin{definition}
Let $\chi_M$ be a slicing on $M$, a \emph{compatible slicing} is a diffeomorphism
\[
\begin{array}{rccl}
\chi_E\colon \bbR\times E_{|\Sigma}\to E
\end{array}
\]
such that the following diagram is commutative
\[
\xymatrix{
\bbR\times E_{|\Sigma} \ar[rr]^{\chi_E} \ar[dd]&& E\ar[dd]\\
\\
\bbR\times \Sigma\ar[rr]^{\chi_M} && M
}
\]
where the vertical arrows are the bundle projections.
\end{definition}

\begin{definition} Consider the vector field $\ppt\in\mathfrak{X}(\bbR\times E_{|\Sigma})$ constructed following the procedure introduced in the previous section to define $\xi_M$, then the vector field $\xi_E=(\chi_E)_*(\ppt)$ is called the \emph{generator} of $\chi_E$. Notice that this vector field projects onto $\xi_M$ defined in \ref{def:generator}.
\end{definition}

\begin{remark} Remember that due to the slicing on $M$ and the volume form $\eta_{\Sigma}$ on $\Sigma$ we can construct the volume form on $M$ given by $dt\wedge \eta_{\Sigma}$. We are making some abuse of notation using $dt$ to denote $\chi_M^*dt$.

\end{remark}

Observe that a compatible slicing induces a trivialization on $J^1\pi^\dag$ by pullback (we are now thinking about $J^1\pi^\dag$ as a bundle of forms). So we have a diffeomorphism
\[
\chi_{J^1\pi^\dag}\colon \bbR\times (J^1\pi^\dag)_{|\Sigma}\to J^1\pi^\dag.
\]
\begin{definition}
The \emph{generator of} $\chi_{J^1\pi^\dag}$ is the vector field
defined by $\xi_{J^1\pi^\dag}=(\chi_{J^1\pi^\dag})_*(\ppt)$, where $\ppt\in\mathfrak{X}(\bbR\times (J^1\pi^\dag)_{|\Sigma})$ is constructed as in the definition of $\xi_E$ and $\xi_M$.
\end{definition}

\begin{definition}
We define the \emph{Cauchy data space}, and denote it by $\FSection{J^1\pi}$, as the set of embeddings
\[
\begin{array}{ll}
\FSection{J^1\pi}=&\{\sigma_{J^1\pi}\colon \Sigma\to J^1\pi\ \textrm{ such that there exists, $\phi\in\Gamma(\pi)$ }
\\ \noalign{\medskip}
&\textrm{satisfying }\sigma_{J^1\pi}=(j^1\phi)\circ\lambda, \textrm{ where }\lambda=\pi\circ \sigma_{J^1\pi}\in\FSection{M}\}
\end{array}
\]
\end{definition}

Using the extended and reduced Legendre transforms (see \eqref{eq:legendre.transform.ext} and \eqref{eq:legendre.transform.red}) we can induce the maps
\[
\begin{array}{rccl}
\FSection{Leg}_{\mathcal{L}}\colon &\FSection{J^1\pi}&\To & \FSection{J^1\pi^\dag}\\ \noalign{\medskip}
&\sigma_{J^1\pi}&\to&\FSection{Leg}_{\mathcal{L}}(\sigma_{J^1\pi})=Leg_{\mathcal{L}}\circ\sigma_{J^1\pi}
\end{array}
\]
and

\[
\begin{array}{rccl}
\FSection{leg}_{\mathcal{L}}\colon &\FSection{J^1\pi}&\To &\FSection{ J^1\pi^\circ}\\ \noalign{\medskip}
&\sigma_{J^1\pi}&\to&\FSection{leg}_{\mathcal{L}}(\sigma_{J^1\pi})=Leg_{\mathcal{L}}\circ\sigma_{J^1\pi}
\end{array}
\]
using the procedure developed in Remark \ref{rmk:composition}.

We introduce now the phase space $\cota$ and relate it with the
previously defined space $\FSection{J^1\pi^\circ}$. Recall that for each $t\in\bbR$ we have the bundle given by restriction to $\Sigma_t$ that we described in \ref{definition:restriction}. Remember that we used the notation
\[
\pi_t\colon E_t\to \Sigma_t
\]
where $E_t=E_{|\Sigma_t}$ and $\pi_t=\pi_{|\Sigma_t}$ and in the same
way we have the analogous restrictions for all bundles involved in our constructions.

\begin{remark}\label{rmk:identification} The space of sections of each of these bundles is denoted in \cite{GIMMsyII} by $\mathcal{E}_t$, i.e., $\mathcal{E}_t=\Gamma(\pi_t)$. The points of $\FSection{E}$ can be identified with points in $\cup_{t\in\bbR}\mathcal{E}_t$. Let $\sigma_E\in\FSection{E}$ and $\FSection{\pi}(\sigma_E)=t_0$, then we can consider $\sigma_E\circ(\chi_M)_{t_0}^{-1}\in\mathcal{E}_{t_0}$, where $(\chi_M)_{t_0}^{-1}$ is the inverse of the restriction to its image of $(\chi_M)_{t_0}$ as introduced previously. In that way, we have a bijection that allows us to identify $\FSection{E}=\cup_{t\in\bbR}\mathcal{E}_t$. We assume that the spaces $\mathcal{E}_t$ are infinite dimensional manifolds modeled on the corresponding functional space.
\end{remark}

In the same way
\begin{equation}
\begin{array}{lr}
(\pi_ 1^\dag)_t\colon J^1\pi^\dag_t\to \Sigma_t;
&
(\pi_1^\circ)_t\colon J^1\pi^\circ_t\to \Sigma_t,
\end{array}
\end{equation}
where
\[
\begin{array}{lr}
J^1\pi^\circ_t=J^1\pi^\circ_{|\Sigma_t}, &J^1\pi^\dag_t=J^1\pi^\dag_{|\Sigma_t}, \\ \noalign{\medskip}
 (\pi_ 1^\circ)_t=(\pi_ 1^\circ)_{|\Sigma_t}, &(\pi_1^\dag)_t=(\pi_1^\dag)_{|\Sigma_t}.
\end{array}
\]

\begin{remark}
The same procedure stated in Remark \ref{rmk:identification} can be applied here on this spaces in an obvious way.
\end{remark}

For a fixed $t$, taking a curve in $\mathcal{E}_t$ it is easy to see that the
\emph{tangent vectors} of this manifold at a point $\sigma_E$ are
given by a section $V$ of the bundle
$\tau_{\Vert}^t\colon\Vert(\pi_t)\to \Sigma_t$ ($\tau_{\Vert}^t$ is
the natural projection), such that $\sigma_E=\tau_{\Vert}\circ V$, that is
\[
T_{\sigma_E}\mathcal{E}_t=\{V\in\Gamma(\tau_{\Vert}) \textrm{, such that } \sigma_E=\tau_{\Vert}\circ V\}.
\]
So the \emph{tangent bundle} is just \[T\mathcal{E}_t=\bigcup_{\sigma_E\in\mathcal{E}_t}T_{\sigma_E}\mathcal{E}_t.\]

We proceed now to introduce the dual space of $T\mathcal{E}_t$. In order to do that, we need to introduce the dual of the bundle $\tau_{\Vert}\colon \Vert(\pi_t)\to \Sigma_t$, which we denote by
\[
\pi_{\Vert}\colon \Vert^*(\pi_t)\to E_t.
\]

The tensor product of the bundles $\pi_{\Vert}\colon \Vert^*(\pi_t)\to E_t$ and $\pi_t^*(\Lambda^m\Sigma_t)\to E_t$, which we refer to $\pi_{\otimes}\colon \Vert^*(\pi_t)\otimes\pi_t^*(\Lambda^m\Sigma_t)\to E_t$ is the space whose sections will give us the dual elements of the tangent vectors.

\begin{definition}[ref. \cite{GIMMsyII}, pag. 72] The \emph{smooth cotangent space to $\mathcal{E}_t$ at a point} $\sigma_E\in \mathcal{E}_t$ is
\[
T^*_{\sigma_E}\mathcal{E}_t=\{\lambda\colon \Sigma\to V^*\pi_t\otimes \Lambda^n\Sigma_t \textrm{ such that }\pi_{\otimes}\circ\lambda=\sigma_E\}.
\]
\end{definition}

\begin{definition}
The \emph{smooth cotangent bundle} is
\[T^*\mathcal{E}_t=\bigcup_{\sigma_E\in\mathcal{E}_t}T^*_{\sigma_E}\mathcal{E}_t.\]
\end{definition}

There is a natural pairing between these two spaces. If $V\in T_{\sigma_E}\mathcal{E}_t$ and $\lambda\in T^*_{\sigma_E}\mathcal{E}_t$ the pairing is given locally by
\[
\int_{\Sigma_t}\lambda(V)
\]

Given coordinates adapted to the bundle $\pi_t\colon E_t\to \Sigma_t$, $(x^i,u^\alpha)$, $1\leq i\leq m$, $1\leq \alpha \leq n$, local coordinates in the space $T\mathcal{E}_t$ are given by
\[
(u^\alpha(\cdot{}),\dot{u}^\alpha(\cdot{}))\to \dot{u}^\alpha(\cdot{})\pp[]{u^\alpha}.
\]
Then we have the corresponding coordinates in $T^*\mathcal{E}_t$
\[
(u^\alpha(\cdot{}),\pi_\alpha(\cdot{}))\to \pi_\alpha(\cdot{})du^\alpha\otimes \dmx
\]
where $\dmx=dx^1\wedge\ldots dx^m$. Similarly to the case in \eqref{coordinates:base}, \eqref{coordinates:reduced} and \eqref{coordinates:extended} the $u^\alpha(\cdot{})$, $\dot{u}^\alpha(\cdot{})$ and $\pi_\alpha(\cdot{})$ are functions that depend on the variables $(x^1,\ldots,x^m)$ and that belong to the chosen functional space.

Now, for each $t\in \bbR$ we have the maps defined by
\[
\begin{array}{rcclccl}
R_t\colon &\FSection{J^1\pi^\dag_t}&\To & T^*\mathcal{E}_t \\ \noalign{\medskip}
&\sigma_{J^1\pi^\dag}&\to& R_t(\sigma_{J^1\pi^\dag})\colon &T \mathcal{E}_t&\To &\bbR\\
 &&&& V&\to &R_t(\sigma_{J^1\pi^\dag})(V)=\displaystyle\int_{\Sigma_t}\phi^*(i_V\sigma_{J^1\pi^\dag})
\end{array}
\]
where $\phi=\nu\circ\sigma_{J^1\pi^\dag}\circ (\chi_M)_{t}^{-1}$ and $(\chi_M)_{t}^{-1}$ should be understood in the sense of Remark \ref{rmk:identification}, and the map

\[
\begin{array}{rcclccl}
R^\circ_t\colon &\FSection{J^1\pi^\circ_t}&\To& T^*\mathcal{E}_t \\\noalign{\medskip}
&\sigma_{J^1\pi^\circ}&\to& R^\circ_t(\sigma_{J^1\pi^\circ})\colon &T \mathcal{E}_t&\To &\bbR\\
 &&&& V&\to &R^\circ_t(\sigma_{J^1\pi^\circ})(V)=\displaystyle\int_{\Sigma_t}\phi^*(i_V\sigma_{J^1\pi^\circ})
\end{array}
\]
where $\phi=\nu^\circ\circ\sigma_{J^1\pi^\circ}\circ (\chi_M)_{t}^{-1}$. Notice that the contraction $i_V\sigma$ is well defined since $V$ is a vertical vector field of the bundle $\pi_t$ . In local coordinates \[R_t(u^\alpha(\cdot{}),p(\cdot{}),p^t_\alpha(\cdot{}),p_\alpha^i(\cdot{}))=(u^\alpha(\cdot{}),\pi_\alpha=p^t_\alpha(\cdot{})).\]

\begin{definition}[ref. \cite{GIMMsyII}, pag. 93]
For each $r\in\bbR$, the \emph{instantaneous Hamiltonian function} is the function
\[
\begin{array}{rccl}
\mathfrak{H}_t\colon &T^*\mathcal{E}_t&\To& \bbR \\ \noalign{\medskip}
&\lambda&\to & \mathfrak{H}_t(\lambda)=-\int_{\Sigma_t}\sigma^*(i_{\xi_{J^1 \pi^\dag}}\Theta)
\end{array}
\]
where $\sigma$ denotes any element in $\mathop{Im}(\FSection{leg}_{\mathcal{L}})\cap (R_t)^{-1}(\lambda)$.
\end{definition}

 Notice that in coordinates, if $\lambda=(u^\alpha(\cdot{}),\pi_\alpha(\cdot{}))$, that means that there exists a point $\sigma_{J^1\pi}\in\FSection{J^1\pi}$ that locally reads $(t,u^\alpha(\cdot{}),u^\alpha_i(\cdot{}),u_0^\alpha(\cdot{}))$ and such that
\[
\pp[L]{u^\alpha_0}(t,x^i,u^\alpha(x^i),u^\alpha_i(x^i),u_0^\alpha(x^i))=\pi_\alpha(x^i)
\quad \textrm{ for all } (x^i)
\]
thus
\[
\begin{array}{l}
-\displaystyle\int_{\Sigma_t}\lambda^*(i_{\xi_{J ^1\pi^\dag}}\Theta)
\\ \noalign{\medskip}

=-\displaystyle\int_{\Sigma_t}\lambda^*\Big(i_{\ppt}\Big(\displaystyle\pp[L]{u^\alpha_i}du^\alpha\wedge d^mx_i+\pp[L]{u^\alpha_0}du^\alpha\wedge d^mx_0+(L-\pp[L]{u^\alpha_i}u^\alpha_i-\pp[L]{u^\alpha_0}u^\alpha_0)d^{m+1}x\Big)\Big)
\\ \noalign{\medskip}

=-\displaystyle\int_{\Sigma_t}\lambda^*\Big(-\displaystyle\pp[L]{u^\alpha_i}du^\alpha\wedge d^{(m-1)}x_{i0}+(L-\pp[L]{u^\alpha_i}u^\alpha_i-\pp[L]{u^\alpha_0}u^\alpha_0)d^mx_0\Big)
\\ \noalign{\medskip}

=-\displaystyle\int_{\Sigma_t}(\displaystyle\pp[L]{u^\alpha_i}\pp[u^\alpha]{x^i}+(L-\pp[L]{u^\alpha_i}\pp[u^\alpha]{x^i}-\pp[L]{u^\alpha_0}\pp[u^\alpha]{x^0}))d^mx_0
\\ \noalign{\medskip}

=\displaystyle\int_{\Sigma_t}(-L+\displaystyle\pp[L]{u^\alpha_0}\pp[u^\alpha]{x^0})d^mx_0.
\end{array}
\]

Now we denote by $\cota$, the bundle over $\FSection{M}\equiv \bbR$ such that, for each $t\in\bbR$ the fiber is $T^*\mathcal{E}_t$. We use
\[
\pi_{\cota}\colon \cota\to \bbR
\]
to denote the projection onto $\bbR$. Notice that we have the following equality of sets $\cota=\bigcup_{t\in\bbR}T^*\mathcal{E}_t$.

Local coordinates in this bundle adapted to the fibration $\pi_{\cota}$ are
\[
(t,u^\alpha(\cdot{}),\pi_\alpha(\cdot{}))\to \pi_\alpha(\cdot{})du^\alpha\otimes \dmx\in T^*\mathcal{E}_t.
\]
where we also assume that $\xi_M$ is given by $\ppt$ in this coordinates.
\begin{remark}

Every tangent vector $X\in T_\lambda\cota$
can be locally written as
\[
X=k\ppt+X_ {u^{\alpha}}\pp[]{u^{\alpha}}+X_{\pi_ {\alpha}}\pp[]{\pi_ {\alpha}}.
\]
\end{remark}

On $T^*\mathcal{E}^t$ there is a form $\omega$ given in local coordinates by
\[
\int_{\Sigma_t}du^\alpha\wedge d\pi_\alpha\otimes \dmx
\]
we explain now that expression. Given two tangent vectors $X,Y\in
T_{\lambda}\cota$ such that in adapted coordinates
\begin{equation}
\begin{array}{l}
X=X_ t\displaystyle\ppt+X_ {u^{\alpha}}\pp[]{u^{\alpha}}+X_{\pi_
 {\alpha}}\pp[]{\pi_ {\alpha}}\\ \noalign{\medskip}

Y=Y_t\displaystyle\ppt+Y_ {u^{\alpha}}\pp[]{u^{\alpha}}+Y_{\pi_ {\alpha}}\pp[]{\pi_ {\alpha}}
\end{array}
\end{equation}
then
\[
\omega(X,Y)=\int_{\Sigma_t} (X_ {u^{\alpha}}Y_{\pi_
 {\alpha}} -X_{\pi_
 {\alpha}}Y_ {u^{\alpha}})\dmx
\]

\begin{remark} This form is obtained gluing together the canonical
 symplectic forms of the cotangent bundles
 $T^*\mathcal{E}_t$. Detailed information about this process can be
 seen in ref \cite{GIMMsyII}, pag. 103.
\end{remark}

\begin{definition}
We define the \emph{Hamiltonian function},
$\mathfrak{H}\colon\cota\to \bbR$ satisfying that for $\lambda$ such that
$\pi_{\cota}(\lambda)=t$ is defined by $\mathfrak{H}(\lambda)=\mathfrak{H}_t(\lambda)$.
\end{definition}

We construct the $2$--form $\omega+d\mathfrak{H}\wedge dt$ on $\cota$.

\begin{definition}A section $c(t)$ of the bundle $\cota$ is called a \emph{dynamical trajectory} if
\[
i_{\dot{c}(t)}(\omega+ d\mathfrak{H}\wedge dt)=0.
\]
Notice that we are using that $c$ is a curve and
denoting by $\dot{c}(t)$ its derivative at time $t$.
\end{definition}

We proceed now, in the same way we did in Theorem
\ref{th:correspondence}, to show the relation between dynamical
trajectories and solutions of the Euler-Lagrange equations. Let $\phi$
be a section of $\pi$ and $j^1\phi$ its first jet bundle. Set the
section of $\pi_1^\dag$ given by
$\sigma_{J^1\pi^\dag}=Leg_{\mathcal{L}}\circ j^1\phi$ and construct the
curve $c:\bbR\to \cota$
\[
c(t)=R_t((\sigma_{J^1\pi^\dag})_{|\Sigma_t}).
\]

In the reference \cite{GIMMsyII} it is shown how to relate the Euler-Lagrange equations with dynamical trajectories.

\begin{proposition}[ref. \cite{GIMMsyII}, pag. 105]\label{th:correspondence2}
The section $\phi$ satisfies the Euler-Lagrange equations if and only if $c(t)$ is a dynamical trajectory.
\end{proposition}

The result below relates the dynamics on the manifold $J^1\pi^\circ$ with the dynamics on $\cota$. We introduce the map $R$ which results from gluing the maps $R_t$
\[
\begin{array}{rccl}
R\colon & \FSection{J^1\pi^\circ} &\To & \cota\\ \noalign{\medskip}
& \sigma_{J ^1\pi^\circ}&\to &R_{t}(\sigma_{J ^1\pi^\circ})
\end{array}
\]
where $t=\FSection{\pi_1^\circ}(\sigma_{J ^1\pi^\circ})$.

\begin{proposition}\label{proposition:last} We have ${R}^*(\omega+d\mathfrak{H}\wedge dt)(\sigma_{J^1\pi^\circ})=\FSection{\Omega}_h(\sigma_{J^1\pi^\circ})$ for all $\sigma_{J^1\pi^\circ}\in$ Im$(\FSection{leg_{\mathcal{L}}})$.
\end{proposition}
\proof
The proof will be done in local coordinates. Let $\FSection{X}$ and
$\FSection{Y}$ be vectors in $T_{\sigma_{J^1\pi^\circ}}$.
\[
\begin{array}{l}
X=X_t\displaystyle\ppt+X_ {u^{\alpha}}\pp[]{u^{\alpha}}+X_{p^t_
 {\alpha}}\pp[]{p^t_
 {\alpha}}+X_{p^i_
 {\alpha}}\pp[]{p^i_
 {\alpha}}\\ \noalign{\medskip}
Y=Y_t\displaystyle\ppt+X_ {u^{\alpha}}\pp[]{u^{\alpha}}+Y_{p^t_
 {\alpha}}\pp[]{p^t_
 {\alpha}}+Y_{p^i_
 {\alpha}}\pp[]{p^i_
 {\alpha}}.
\end{array}
\]

 By the definition of $\FSection{\Omega}_h$ we have
\[
\FSection{\Omega}_h(\FSection{X},\FSection{Y})=\int_{\Sigma_t}(\sigma_{J ^1\pi^\circ})^*i_{\FSection{Y}} i_{\FSection{X}}\Big(-dp_\alpha^t\wedge du^\alpha\wedge d^mx_i-dp_\alpha^i\wedge du^\alpha\wedge d^mx_0+dH\wedge d^{m+1}x\Big)
\]

Recall that $H=\displaystyle\pp[L]{u^\alpha_i}u^\alpha_i+\displaystyle\pp[L]{u^\alpha_0}u^\alpha_0-L$. Using this we can split
\[
\begin{array}{l}
\displaystyle\int_{\Sigma_t}(\sigma_{J ^1\pi^\circ})^*i_{\FSection{Y}} i_{\FSection{X}}\Big(-dp_\alpha^t\wedge du^\alpha\wedge d^mx_0-dp_\alpha^i\wedge du^\alpha\wedge \dmx+dH\wedge \dmpx\Big)\\

=\underbrace{\int_{\Sigma_t}(\sigma_{J ^1\pi^\circ})^*i_{\FSection{Y}} i_{\FSection{X}}\Big(-dp_\alpha^t\wedge du^\alpha\wedge d^mx_0+d(\pp[L]{u^\alpha_0}u^\alpha_0-L)\dmpx}_{i)}\Big)\\

+\underbrace{\int_{\Sigma_t}(\sigma_{J ^1\pi^\circ})^*i_{\FSection{Y}} i_{\FSection{X}}\Big(-dp_\alpha^i\wedge du^\alpha\wedge d^mx_i+d(\pp[L]{u^\alpha_i}u^\alpha_i)\dmpx}_{ii)}\Big)
\end{array}
\]

The first part is easily seen to be equal to $ (\omega+d\mathfrak{H}\wedge dt)(R( \sigma_{J ^1\pi^\circ}))(TR(\FSection{X},\FSection{Y}))$. We will show that $ii)$ vanishes. Taking into account that
\[
\begin{array}{l}
-dp_\alpha^i\wedge du^\alpha\wedge d^mx_i+d(\displaystyle\pp[L]{u^\alpha_i}u^\alpha_i)\dmpx
\\ \noalign{\medskip}
=d\Big(-p_\alpha^i du^\alpha\wedge d^mx_i+(\displaystyle\pp[L]{u^\alpha_i}u^\alpha_i)\dmpx\Big)
\end{array}
\]
and using Lemma \ref{lemma:differential} we are done if we show
that $\FSection{\beta}=0$, where $\beta=-p_\alpha^idu^\alpha\wedge
d^mx_i+(\pp[L]{u^\alpha_i}u^\alpha_i)d^{m+1}x$. We show now that this form vanishes for every tangent vector on $J^1\pi^\circ$:

Recall that a tangent vector, $\FSection{\xi}\in T_{\sigma_{J
 ^1\pi^\circ}}\FSection{J^1\pi^\circ}$ is given locally by \[\FSection{\xi}=k\ppt+\xi_{u^\alpha}(\cdot{})\pp[]{ u^\alpha} +\xi_{p_\alpha^t}(\cdot{})\pp[]{p_\alpha^t}+\xi_{p_\alpha^i}(\cdot{})\pp[]{p_\alpha^i}\]
and so
\[
\FSection{\beta}(\FSection{\xi})=\FSection{\beta}(k\ppt+\xi_{u^\alpha}\pp[]{ u^\alpha})=\FSection{\beta}(k\ppt)+\FSection{\beta}(\xi_{u^\alpha}\pp[]{ u^\alpha})
\]
since the form $\beta$ does not contract with $\displaystyle\pp[]{p_\alpha^t}$ and $\displaystyle\pp[]{p_\alpha^i}$.

First of all notice that
\[
\begin{array}{l}
\FSection{\beta}(\sigma_{J^1\pi^\circ})(\pp[]{
 t})=\displaystyle\int_{\Sigma_t}(\sigma_{J^1\pi^\circ})^*(i_{\frac{\partial}{\partial
 t}}(-p_\alpha^i du^\alpha\wedge
d^m x_i+\displaystyle\pp[L]{u^\alpha_i}u^\alpha_i d^{m+1}x))
\\ \noalign{\medskip}

=\displaystyle\int_{\Sigma_t}(\sigma_{J^1\pi^\circ})^*(p_\alpha^i du^\alpha\wedge d^{m-1}x_{i0}+\displaystyle\pp[L]{u^\alpha_i}u^\alpha_i d^mx_0)
\\ \noalign{\medskip}

=\displaystyle\int_{\Sigma_t}(\sigma_{J^1\pi^\circ})^*(p_\alpha^i du^\alpha\wedge d^{m-1}x_{i0}+p_\alpha^iu^\alpha_id^mx_0)
\\ \noalign{\medskip}

=\displaystyle\int_{\Sigma_t}\left(-((\sigma_{J^1\pi^\circ})_\alpha^i) \displaystyle\pp[(\sigma_{J^1\pi^\circ})^\alpha]{x^i}+((\sigma_{J^1\pi^\circ})_\alpha^i)\pp[(\sigma_{J^1\pi^\circ})^\alpha]{x^i} \right)d^mx_0=0
\end{array}
\]
where $(\sigma_{J^1\pi^\circ})_\alpha^i=p_\alpha^i\circ\sigma_{J^1\pi^\circ}$ and $(\sigma_{J^1\pi^\circ})^\alpha=u^\alpha\circ \sigma_{J^1\pi^\circ}^\alpha$ .

On the other hand
\[\FSection{\beta}(\xi_{u^\alpha}\pp[]{ u^\alpha})
=\int_{\Sigma_t}((\sigma_{J^1\pi^\circ})_\alpha^i)\xi_\alpha
(\sigma_{J^1\pi^\circ}^*(-dt\wedge d^mx_i))=0,\] because $(\sigma_{J^1\pi^\circ})^*dt=0$.
\qed

With this result at hand, we can now induce a Hamilton-Jacobi theory
on the space $\cota$ following the same pattern as in Section \ref{sec:cft:hamilton-jacobi:cauchy}.
Assume now that we have $\gamma$ satisfying the Hamilton-Jacobi equation. We can define
\[
\begin{array}{rccl}
\hat{\gamma}\colon & \FSection{E}&\To &\cota\\ \noalign{\medskip}
& \sigma_E&\to &\hat{\gamma}(\lambda)=R\circ\mu\circ\gamma\circ\sigma_E
\end{array}
\]

\begin{remark}\label{rmk:??}
We are assuming that $R\circ \FSection{leg_{\mathcal{L}}}$ is a bijection between $\FSection{J^1\pi}$ and $\cota$, so in particular for any $\lambda\in \cota$ there exists $\sigma_{J^1\pi^\circ}\in R^{-1}(\lambda)\cap$Im$(leg_{\mathcal{L}})$.
\end{remark}

As an immediate consequence of the previous theorem we conclude the following proposition.
\begin{proposition} Under the previous assumptions, we have:
\begin{enumerate}
\item $\hat{\gamma}^*(\omega+ d\mathfrak{H}\wedge dt)=0$
\item $i_{T\hat{\gamma} (\sigma_E)(X^{\hor^\gamma})}(\omega+ d\mathfrak{H}\wedge dt)=0$ for all $\sigma_E\in \FSection{E}$ which is an integral submanifold of the connection $\hor^\gamma_{\FSection{\pi}(\sigma_E)}$.
\end{enumerate}
\end{proposition}
\proof

The proof is a consequence of Proposition \ref{proposition:last} and Remark \ref{rmk:??}.

\qed

We want to finish this section setting some useful
identifications. Using the vector field $\xi_{J^1\pi^\circ}$ the space
$\cota$ can be identified with $\bbR\times T^*\mathcal{E}_t$ for a
fixed $t$. Under this identification the Hamiltonian function becomes
a function on $\bbR\times T^*\mathcal{E}_t$ and the form $\omega$
becomes the canonical form on the cotangent bundle
$\omega_{\mathcal{E}_t}$. It is worth to notice that this becomes not
the usual spaces where the classical Hamilton-Jacobi theory is developed. 

\nocite{*}

\appendix
\section{Ehresmann Connections}

Let $\pi : E \longrightarrow M$ be a fibred bundle (that is, $\pi$ is a surjective submersion).
Denote by $VE$ the vertical bundle defined by $\ker \pi$ which is a vector sub-bundle of $TE \longrightarrow E$.

\begin{definition} An Ehresmann connection in $\pi :E \longrightarrow M$ is a distribution $\mathbf{H}$ on $E$ which is complementary to the vertical bundle, say
\begin{equation}\label{horizontal}
TE = \mathbf{H} \oplus VE
\end{equation}
$\mathbf{H}$ is called the horizontal distribution.
\end{definition}

Given a connection $\mathbf{H}$ in $\pi : E \longrightarrow M$ we have two complementary projectors:
\begin{eqnarray*}
&& h : TE \longrightarrow \mathbf{H} \\
&& v : TE \longrightarrow VE
\end{eqnarray*}
$h$ and $v$ are called the horizontal and vertical projectors, respectively.

Obviously, we have $\mathbf{H} = \mathop{Im}(h)$ and $VE = \mathop{Im}(v)$. Consequently, any tangent vector $X \in T_eE$ can be decomposed in its 
horizontal and vertical parts, say
$$
X = hX + vX
$$
In addition, given a tangent vector $Y \in T_xM$, there exists a unique tangent vector $X$ at any point of the fiber over $x$,
say $e \in \pi^{-1}(x)$ such that $X$ is horizontal and projects onto $Y$; $X$ is called the horizontal lift of $Y$ to $e$.

The curvature of a connection $\mathbf{H}$ (or $h$ with some abuse of notation) can be defined as the Schouten-Nijenhuis bracket
$$
R = - \frac{1}{2} [h, h]
$$
such that $\mathbf{H}$ is flat if and only if $R = 0$.

A connection $\mathbf{H}$ should not be flat in general; let us introduce the notion of integral section.

\begin{definition}
A section $\gamma : M \longrightarrow E$ is called an integrable section of $\mathbf{H}$ if $\gamma(M)$ is an integral submanifold of the horizontal distribution.
The connection $\mathbf{H}$ is integrable if and only if there are integral sections passing through any point of $E$.
\end{definition}

Therefore, we easily have the following result.

\begin{theorem}
A connection $\mathbf{H}$ is integrable if and only if it is flat.
\end{theorem}
Indeed, $R = 0$ which is just the condition for the integrability of the distribution $\mathbf{H}$.

\bibliographystyle{plain}
\bibliography{hjcft}

\end{document}